# Recent Advances in mm-Wave and Sub-THz/THz Oscillators for FutureG Technologies

**BAKTASH BEHMANESH*[1], AND AHMAD REZVANITABAR*[2], (Member, IEEE)**
[1]Department of Electrical and Information Technology, Lund University, Lund, 221 00, Sweden (e-mail: baktash.behmanesh@eit.lth.se)
[2]School of Electrical and Computer Engineering, Georgia Institute of Technology, Atlanta, GA 30332, USA (e-mail: ahmadrezvanitabar@gmail.com)

* The authors contributed equally to this work.

**ABSTRACT** This paper provides a concise yet comprehensive review of recent advancements in millimeter-wave (mm-wave) oscillators below 100 GHz and sub-terahertz (sub-THz/THz) oscillators above 100 GHz for next-generation computing and communication systems, including 5G, 6G, and beyond. Various design approaches, including CMOS, SiGe, and III-V semiconductor technologies, are explored in terms of performance metrics such as phase noise, output power, efficiency, frequency tunability, and stability. The review highlights key challenges in achieving high-performance and reliable oscillator designs while discussing emerging techniques for performance enhancement. By evaluating recent design trends, this work aims to offer valuable insights and design guidelines that facilitate the development of robust mm-wave and sub-THz/THz oscillators for future communication, computing, and sensing applications.

## I. INTRODUCTION

**T**HE widespread integration of communication technologies in modern society necessitates the exploration of new frequency bands within the spectrum. These bands enable higher data rates, which are essential for supporting bandwidth-intensive applications such as video streaming and VR/XR. Achieving these higher data rates often requires larger channel bandwidths, which are predominantly available in previously unutilized mm-wave and sub-THz/THz frequency ranges. A fundamental component in nearly all wireless and wireline communication transceivers is the frequency generation block, responsible for producing a stable oscillation used in key functions such as up-conversion in transmitters, down-conversion in receivers, and signal sampling in data converters. Advances in semiconductor process technology, particularly with higher transition frequencies ($f_T$), have provided new opportunities for designing oscillators operating at mm-wave and sub-THz frequencies. These oscillators can be realized through either direct frequency synthesis or by leveraging frequency multipliers to shift lower-frequency signals into higher bands.

One of the critical metrics for evaluating the effectiveness of a frequency generation block is its spectral purity, often quantified by the phase noise (PN). The oscillator's PN is frequently attributed to the thermal or 1/f noise of various components within the circuit, which is then up-converted to the oscillation frequency. The PN has a detrimental effect on the signal-to-noise ratio (SNR) across diverse communication systems, including carrier-based radar systems, wireless transceivers impacted by reciprocal mixing, and sampled-data systems affected by clock jitter [1]. The reported PN has shown significant improvement in recent years, driven by a growing push toward achieving $-120\,\mathrm{dBc/Hz}$ at 1 MHz offset from the central frequency as shown in Fig. 1. This trend is fueled by modern applications that demand higher performance under increasingly stringent system-level requirements, such as throughput exceeding Gbps in wideband mm-wave communication links, reliable gigabit backhaul, and sub-cm range resolution, accurate velocity extraction, and clean target tracking in FMCW automotive radars. In sensing applications where the information is encoded as phase, frequency, or timing modulation, oscillator PN directly limits resolution by adding random phase/frequency fluctuations. In coherent sensing systems, this appears as timing jitter and reduced phase coherence, degrading range/phase estimates. In heterodyne or PLL-based sensing, the PN is translated to baseband by mixing and becomes indistinguishable from true sensor variations. Close-in phase noise is especially critical when sensor information resides near the carrier, as noise sidebands raise the effective noise floor and reduce minimum detectable signal; thus, oscillator phase noise often limits sensitivity, dynamic range, and required averaging time in integrated sensing front-ends. Consequently, there is a continuous pursuit among designers to devise innovative approaches for mitigating the PN in oscillators. This effort, in turn, enhances the performance of communication systems, enabling the implementation of higher data rates through more advanced modulation schemes.

Research on 6G is accelerating while innovative technologies like distributed massive-MIMO face challenges with phase synchronization and drift among multiple antennas [2]. These issues underscore the importance of spectral purity in frequency generation blocks, especially as they extend into sub-THz frequency bands. With the rapid growth of artificial intelligence systems and cloud computing, next-generation wireline transceivers are targeting data rates of $3.2\,\text{TbE}$, necessitating a $448\,\text{Gb/s}$ data rate per lane [3]. These extreme data rates also demand extremely low jitter levels. For instance, a PAM-4 wireline transmitter operating at $226\,\text{Gb/s}$ utilizing a $56\,\text{GHz}$ PLL must maintain an RMS jitter level within a few percent of the symbol period, i.e. $8.93\,\text{ps}$, translating to merely $100\,\text{fs}$ of RMS jitter [4]. Additionally, the widespread deployment of Internet of Things (IoT) devices, including wearables, and the rise of the metaverse concept and edge Artificial Intelligence (AI) require highly energy-efficient signal generation to extend the battery life of portable devices [5]. Furthermore, emerging wireless communication paradigms, such as *Integrated Sensing and Communication (ISAC)*, are driving the next wave of innovation in integrated oscillator design. ISAC systems aim to seamlessly integrate wireless communication with high-precision sensing, enabling applications such as autonomous vehicles, radars, smart healthcare, and advanced imaging and surveillance systems. These dual-purpose systems place unprecedented demands on spectral purity, as phase noise and spurious tones can significantly degrade both communication throughput and sensing accuracy. The evolution towards 6G networks is increasingly centered on the upper mid-band (FR3) spectrum roughly from 7 to 24 GHz, which is widely regarded as the "Golden Band" for achieving an optimal compromise between the extensive coverage of legacy sub-6-GHz networks and the gigabit-level capacity of millimeter-wave frequencies. However, harnessing this spectral potential imposes stringent requirements on local oscillator design, necessitating architectures that can deliver ultra-low phase noise and wide tuning ranges to support high-order modulation schemes and ensure interference-free operation, within these dense frequency bands. Simultaneously, for radar and imaging applications, the spectral purity of the local oscillator directly dictates Doppler resolution and detection sensitivity. Consequently, minimizing phase noise is critical to prevent target masking and ensure high-fidelity sensing in emerging 6G and ISAC systems.

This overview aims to present recent advancements in mm-wave and sub-THz/THz oscillators for 5G, 6G, and beyond, and it is intended as a broad and structured survey, with a focus on categorizing and summarizing key innovations across a large body of literature. The review focuses on IEEE publications from major journals and conferences published in recent years, including the IEEE Journal of Solid-State Circuits (JSSC), International Solid-State Circuits Conference (ISSCC), European Solid-State Circuits Conference (ESSCIRC), IEEE Transactions on Circuits and Systems I and II (TCAS-I and TCAS-II), IEEE Transactions on Microwave Theory and Techniques (TMTT), Custom Integrated Circuits Conference (CICC), and Radio Frequency Integrated Circuits Symposium (RFIC). The remainder of the article is organized as follows. Section II provides a brief introduction to the fundamental concepts in oscillator design. Sections III and IV discuss fundamental-tone and harmonic-tone oscilla-tors, respectively, for mm-wave frequencies below 100 GHz. Section V covers recent developments in sub-THz and THz oscillators operating at frequencies above 100 GHz. Finally, Section VI presents the concluding remarks.

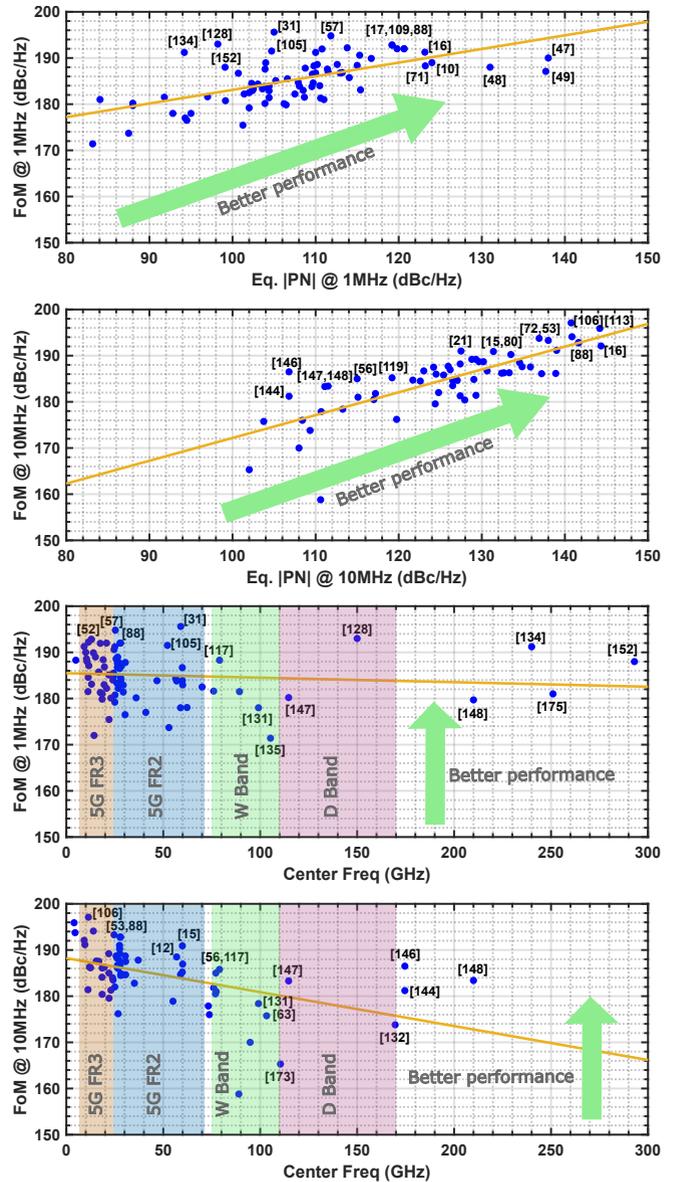

**FIGURE 1.** The reported FoM versus the PN and the center frequency at $1\,\text{MHz}$ and $10\,\text{MHz}$ offset frequencies.

## II. FUNDAMENTAL CONCEPTS OF OSCILLATORS

The most widely employed metric to assess the performance of the oscillators is expressed as the figure of merit (FoM) defined as

$$\text{FoM} = -\mathcal{L}(\Delta\omega) + 20\log_{10}\left(\frac{\omega_0}{\Delta\omega}\right) - 10\log_{10}\left(10^3 \times P\right) \quad (1)$$

where $\mathcal{L}(\Delta\omega)(<0)$ is the PN of the oscillator expressed in dBc/Hz at a certain offset frequency (i.e., $\Delta\omega$) from the frequency of oscillation (i.e., $\omega_0$), and $P$ is the $DC$ power consumption of the oscillator that is traditionally normalized to 1 mW. Clearly, the FoM defined in (1) does not incorporate the tunability of the center frequency, which is an important aspect of the oscillator performance. Therefore, an alternative FoM is frequently employed to include the tuning range (TR) and is defined as

$$\text{FoM}_T = \text{FoM} + 20\log_{10}\left(\frac{TR(\%)}{10}\right) \quad (2)$$

We should mention that the widespread usage of FoM$_T$ is questionable due to the absence of a solid theoretical foundation, unlike FoM as defined in (1) [1]. The reported FoMs in recent years demonstrate significant improvements, pushing toward a gold-level milestone of 200 dBc/Hz [see Fig. 1]. The equation developed by Leeson in [6] is arguably the most renowned method for computing the PN of a simple LC oscillator, expressed as

$$\mathcal{L}(\Delta\omega) = 10\log_{10}\left(F \cdot \frac{2k_B T R}{A_{pk}^2/2}\left[1 + \left(\frac{1}{2Q}\frac{\omega_0}{\Delta\omega}\right)^2\right]\right.$$
$$\left. \cdot \left(1 + \frac{\Delta\omega_{1/f^3}}{\Delta\omega}\right)\right) \quad (3)$$

where $F$ encapsulates the influence of all white noise sources excluding the tank itself. $k_B$ denotes Boltzmann's constant, while $T$ represents the absolute temperature. $R$ symbolizes the equivalent parallel resistance of the LC tank, while $A_{pk}$ is the voltage amplitude of oscillation. The parameter $Q$ characterizes the quality factor of the tank, with $\omega_0$ indicating the oscillation frequency. $\Delta\omega$ is the offset frequency used for PN assessment, while $\Delta\omega_{1/f^3}$ defines the corner frequency for up-conversion of $1/f$ noise near the oscillation frequency.

Leeson's model is a linear, time-invariant approximation of a nonlinear, time-variant oscillator and therefore provides intuition rather than predictive accuracy. In particular, the $1/f^3$ region is introduced empirically and cannot be reliably predicted across topologies. The development of a linear time-variant model for the PN in [7] led to the introduction of the impulse sensitivity function (ISF), which remains a prevalent approach for understanding PN in oscillator circuits. The ISF method's appeal lies in its ability to capture the time-varying nature of PN generation. For instance, injecting a noise impulse at the peak of a sinusoidal signal results in minimal PN degradation, whereas injecting it at zero crossings maximizes PN generation. Hence, any current noise source, $i_n$, must be multiplied by a corresponding ISF weight factor, $\Gamma_{i_n}$, in order to capture its effective contribution to the PN [1], [7]

$$i_{n,eff}(\phi) = i_n(\phi) . \Gamma_{i_n}(\phi) \quad (4)$$

where $\phi = \omega_0 t$ is the phase of the oscillation. Here, $i_n(\phi)$ denotes the device current noise and $\Gamma_{in}(\phi)$ captures the oscillator's phase sensitivity to noise injected at phase $\phi$. Their product represents the portion of the device noise that is effectively converted into the PN. The ISF is often normalized to be dimensionless, and independent of frequency and amplitude, while having a period of $2\pi$. If $i_n$ is a cyclo-stationary noise source, it can be represented as the multiplication of a wide-sense stationary (WSS) noise source with a modulating function that reflects the time variation of the noise process

$$i_{n,eff}(\phi) = i_{n,WSS}(\phi) . \Gamma_{i_n,eff}(\phi) \quad (5)$$

where all time variance is encapsulated in an effective $\Gamma_{i_n,eff}(\phi)$ [1], [7]. The contribution of a stationary or cyclo-stationary white current noise source, $i_n$, to the PN can be shown to be [7], [8]

$$\mathcal{L}(\Delta\omega) = 10\log_{10}\left(\frac{\overline{i_{n,WSS}^2}/\Delta f \Gamma_{i_n,eff,rms}^2}{2C^2 A_{pk}^2 \Delta\omega^2}\right) \quad (6)$$

where $\overline{i_{n,WSS}^2}/\Delta f$ is the power spectral density (PSD) of the $i_{n,WSS}$ and $\Gamma_{i_n,eff,rms}$ is the root-mean-square (rms) value of $\Gamma_{i_n,eff}$. Fig. 1 illustrates the reported FoM versus the PN and the center frequency for the works that have been covered in this study. Some notable designs are explicitly mentioned with the corresponding reference number.

## III. DESIGN TECHNIQUES FOR MM-WAVE FUNDAMENTAL-TONE HARMONIC OSCILLATOR DESIGN

The primary type of oscillators used in the mm-wave band are harmonic oscillators that operate at the fundamental resonance frequency of the resonator. These so-called fundamental tone oscillators offer several advantages over alternative approaches such as harmonic generation, frequency mixing, or multiplication. In addition to their superior power efficiency, their simpler architecture facilitates easier design and reduces integration complexity in high-frequency systems. By eliminating the need for additional stages, fundamental tone oscillators can achieve a lower PN while maintaining better linearity. However, their limited bandwidth and TR pose challenges for wideband applications. Ensuring signal integrity and frequency stability while minimizing PN degradation remains a key focus in oscillator design. To overcome such limitations, recent techniques are discussed in this section. Although some designs employ two or more techniques, each is included in only the most relevant of the following subsections.

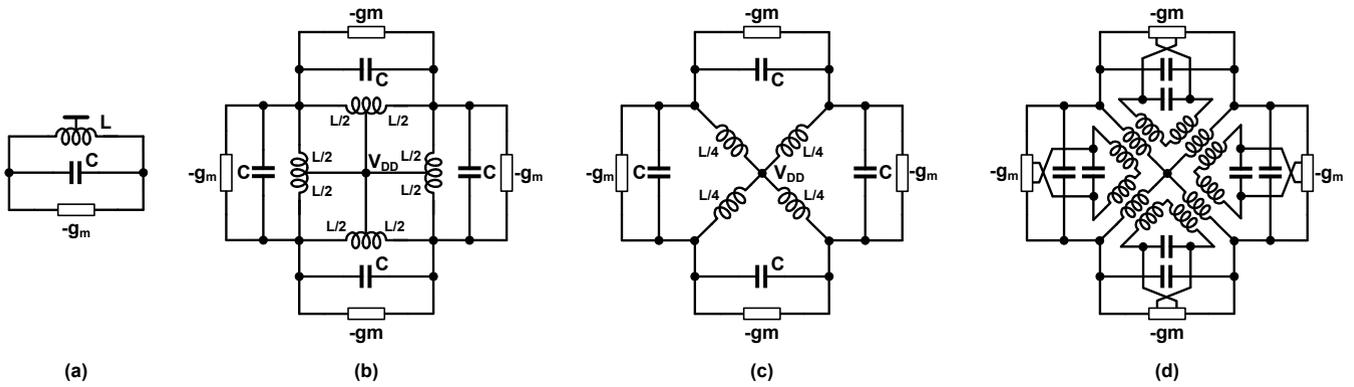

**FIGURE 2.** (a) single core parallel tank VCO (b) quad core VCO (c) inductor sharing to save area (d) addition of secondary windings for class-F operation [9]. A single-core VCO consists of an LC tank in parallel with a negative resistor $-g_m$. Four identical cores connected head-to-head form a quad-core ring-topology VCO. However, its layout is bulky, occupying more than four times the area of a single-core VCO due to unused corner space. To address this, an inductor-sharing technique is introduced to modify the quad-core topology and significantly reduce chip area.

### A. MULTI-CORE TECHNIQUES

A method of achieving significantly low PN levels is the bilateral coupling of N oscillators that leads to a nominal PN reduction of up to $10\log_{10}(N)$. However, the coupling network must be carefully designed to fully take advantage of PN reduction and avoid concurrent multitone oscillations. In [10] (15 GHz with 16% TR, 130 nm BiCMOS), a quad-core bipolar VCO is proposed using a resistive coupling network with optimized resistor values to achieve a stable single-tone oscillation with low PN. In [11] (18.6 to 40.1 GHz, 40 nm CMOS), mixed electric and magnetic coupling between four oscillator cores is proposed to introduce four oscillation modes and significantly boost TR to 73%. In [12] (53.6 to 60.2 GHz, 65 nm CMOS), an XFMR-based mode-rejection-coupled many-core fundamental oscillator is proposed.

In [13] (19.5 GHz, 12.5% TR, 28 nm CMOS), class-C operation is utilized to reduce upconversion of the $1/f$ noise generated by the cross-coupled transistor pair. In [14] (21.6 to 25.5 GHz, 40 nm CMOS), a dual-core VCO with inductive coupling between the cores is presented. Four switches are responsible for switching between odd and even modes of operation, and the ISF function is extracted for various harmonics and the two modes of operation to derive design guidelines for reducing PN level within the operation frequency. In [15] (54.6 to 65.1 GHz, 65 nm CMOS), a 16-core oscillator is presented to reduce PN at mm-wave frequencies. The distinctive feature of this work is the addition of extra synchronization paths between the cores to maintain a low PN penalty induced by the frequency mismatch among various cores. The cores are grouped into four identical groups, each with triple synchronization paths to others, reducing the mismatch due to the asymmetric XFMR layout.

Manipulating second and third harmonics helps reduce wideband flicker noise, achieving a lower $1/f^3$ corner frequency. [16] (6.8 to 11.6 GHz, 65 nm CMOS) presents a quad-core dual-mode class-F VCO operating under standard $V_{DD}$ with no reliability issues. It voids an asymmetrical topology that sacrifices the tank's Q and circular inductors without CM harmonic manipulation, resulting in a large $1/f^3$ corner frequency. The quad-core configuration includes two identical dual-core units placed orthogonally, where even and odd modes depend on the magnetic flux enhancement or cancellation. A T-gate switch controls the relative current direction of the horizontal and vertical units. For CM 2nd harmonic manipulation, the CM return path consists of two independent resonance tanks, and the second tank includes a switchable circular tail inductor routed outside the first oscillator. This not only provides space for adding resistors between cores and preventing the oscillator from latch-up, but also results in a small DM-to-CM coupling factor by minimizing the impact of the circular tail inductor on the DM oscillation.

To improve oscillation performance, CM noise self-cancellation and isolation techniques can be employed in multi-core oscillators. As an example, [17] (11.5 to 14.3 GHz, 65 nm CMOS) presents an enhanced class-F dual-core VCO based on these techniques, where it also uses 8-shaped inductors. To reduce CM noise injected into the VCO from the supply and the ground rails, the CM noise currents of the drain and source coils are in the opposite direction canceling the magnetic field of each other. Furthermore, to isolate the noise coupling path between the drain and gate terminals, the coupling noise from the drain inductor to the gate is oriented in the opposite direction to the induced current in the gate coil. Fig. 2(a)–(d) illustrate the evolution of the proposed VCO topology in [9] (17 to 19 GHz, 65 nm CMOS). The basic single-core VCO comprises an LC tank and a negative resistance element, as shown in Fig. 2(a). Four such cores are arranged in a ring configuration with adjacent outputs connected to form a quad-core VCO in Fig. 2(b). To reduce chip area, an inductor-sharing technique is introduced, enabling a transformation from star to square geometry. This allows inductors to be shared between adjacent cores and reduces total inductance while maintaining the oscillation frequency, as shown in Fig. 2(c). In Fig. 2(d), a secondary winding is added to the shared inductor structure to enable class-F operation, enhancing phase noise performance without increasing layout size. It is worth mentioning that the square geometry could suffer from mode ambiguity, which can be resolved by adding mode rejection resistors between adjacent ports of the XFMR.

Multi-core approaches open room for further performance improvements. For instance, combining mode switching with multi-core approaches offers area efficiency. [18] (18.6 to 40.1 GHz, 40 nm CMOS) proposes a quad-core oscillator with 2D mode switching employing electromagnetic (EM) mixed-coupling for resonance boosting. It includes four reconfigurable resonances while using only two resonant tanks. It retains a symmetrically coupled XFMR for magnetic coupling and coupling capacitors for electrical coupling. In addition, its 2D mode switch array offers quad mode switching without degrading the oscillator's performance in a given mode, thus reducing PN and extending the TR simultaneously. In another multi-core approach, each core can be a stacked oscillator. For example, in [19] (55 to 63 GHz, 28 nm CMOS), a fundamental dual- and quad-core VCO is presented in which each core is based on stacking and magnetic coupling of two NMOS-based resonators. Not only does it provide robust parasitic mode suppression, but it also increases the tank energy to reduce PN and maintain low-voltage swing for enhanced reliability. Also, an implicit CM impedance is accommodated to reduce PN without demanding an extra-large tail inductor.

It is well-known that synchronizing multiple VCO cores reduces the total equivalent inductance, thereby improving PN. [20] (52.4 to 60.4 GHz, 65 nm CMOS) presents a low-PN quad-core fundamental VCO utilizing circular triple-coupled XFMRs, as illustrated in Fig. 3. A key consideration in multi-core VCOs is the sensitivity to frequency mismatches between cores, which can degrade PN or disrupt synchronization. The proposed design addresses this through a robust synchronization mechanism based on inductance reallocation enabled by the triple-coupled XFMR structure. This method effectively compensates for mismatches and has been verified by measurement. The mode-rejection circular XFMR tank offers three primary advantages that contribute to PN improvement. First, the circular topology enhances the Q of small inductors by mitigating inner-edge deconstructive coupling. Second, the triple-coupled XFMR increases the gate-to-source voltage swing of the $g_m$ pair and allows greater flexibility in suppressing undesired oscillation modes. Third, sharing a single XFMR between two cores improves area efficiency. Additional features include resistors at the gate coil center taps for unwanted mode suppression, AC-grounded drain coil center taps that enable an NMOS-only configuration, and support for harmonic impedance shaping through source feedback, eliminating the need for additional tail-filtering XFMRs.

In [21] (26.1 to 29.9 GHz, 65 nm CMOS), a dual-core coupled LC VCO leverages a fully-balanced complementary coupled-Class-C design in which implicit CM resonance at twice the fundamental tone is used to reduce the flicker noise while resonating at both DM and CM. An octave TR quad-core VCO is presented in [22] (17 to 36.4 GHz, 28 nm CMOS), leveraging a quad-mode XFMR-based inductor with a symmetric design. The quad mode inductor has only four design parameters and consists of two inner and four outer XFMRs. The quad mode operation is achieved by controlling the mode switches between the adjacent and distant cores, preventing switch loss, Q degradation, and widening inductive frequency tuning. It adds a decoupling capacitor between the primary and secondary windings of the outer XFMR to reduce the coupling factor and consequently avoid tuning gaps.

Automatic mode tracking in a multi-mode topology is used in [23] (18.5 to 36.5 GHz, 65 nm CMOS) where a quad-core triple-mode VCO leverages mode tracking output buffers using switched inductors. Good PN is achieved by using two identical single-turn coils that are magnetically and symmetrically coupled. The setup allows for three equivalent inductances to be switched by controlling the direction of the current. Additionally, the automatic mode tracking output buffer increases the amplitude and conserves power by utilizing two magnetically coupled inductors. A quad-core circular geometry class-F VCO is presented in [24] (17.9 to 20.7 GHz, 65 nm CMOS). The mode ambiguity is suppressed by adding cross-coupled narrow and thin metal traces in the class-F XFMR secondary loop. The LC filters of the tail nodes are also included in the circular geometry to save chip area.

To summarize, in coupled-resonator integrated LC VCOs, the coupling topology reshapes the oscillator eigenmode spectrum rather than only suppressing unwanted modes. Circular XFMR coupling enforces rotational symmetry and distributed mutual inductance, producing well-defined spatial phase modes whose frequencies and effective losses (Q) differ by mode, often favoring traveling-wave–like operation and enabling multi-phase (and in symmetric cases quadrature) outputs via near-degenerate mode pairs. Mesh topologies add coupling paths and loops that increase modal separation and distribute resonant energy, which can reduce localized loss and improve robustness to mismatch and loading. Thus, circular XFMR and mesh implementations primarily enhance mode stability through modal engineering (frequency splitting and mode-dependent damping), not solely through unwanted-mode suppression.

### B. WAVEFORM SHAPING TECHNIQUES
In [25] (9.9 to 21.9 GHz, 40 nm CMOS), a single-core dual-mode oscillator is presented, where a switch selects the DM or CM operation to enhance the overall TR. In the DM, the circuit acts as a class-$F^{-1}$ VCO with ISF reshaping for improved PN performance. In the CM, the VCO is transformed into a Colpitts oscillator with an enhanced voltage swing to suppress the noise of the transistors. Overall, the circuit achieved an 84% TR without the performance degradation typically associated with multi-core VCOs, due to effects such as destructive magnetic field coupling or mismatch between cores.

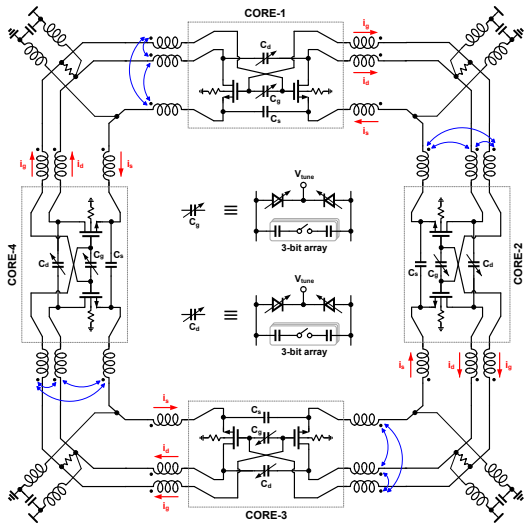

**FIGURE 3.** A low-PN quad-core fundamental VCO with circular triple-coupled XFMRs: it shows more flexibility in suppressing unwanted oscillation modes while avoiding extra area-overhead for tail filtering XFMRs [20].

In [26] (22.5 to 28 GHz and 68 to 84 GHz, 16 nm FinFET), a dual-band frequency generator for radar applications is presented. Here, an oscillator generates both the fundamental and third harmonic frequencies. The fundamental frequency is directly forwarded for phase detection in the ADPLL, while the third harmonic is directed towards a buffer/PA that rejects the fundamental frequency to meet the emission mask requirements. In this way, the 26-GHz signal is intrinsically present in the system without the need to have a dedicated divide-by-three circuit.

In [27] (53.3 to 60.8 GHz, 65 nm CMOS), harmonic positive feedback is utilized in a class-D oscillator to boost the amplitude of the fundamental signal, leading to reduced PN and power consumption. It has been shown that exploiting the transistors operating in the deep triode and using the second harmonic to enhance the gain can improve PN performance. In [28] (9 to 13.34 GHz, 22 nm FDSOI), a complementary class-F oscillator with flicker PN effect reduction utilizing gate-drain phase shift is presented. The phase shift is made tunable by varying the gate-drain capacitance ratio in order to compensate for the detrimental effect of the third-harmonic voltage in a wide frequency range.

A shaped ISF can concentrate noise sensitivity in less critical parts of the oscillation cycle, effectively suppressing the impact of noise sources in key spectral regions. In the work presented in [29] (24.62 to 28.66 GHz, 65 nm CMOS), a multi-resonant 8-port $RLCM$ tank is implemented alongside both NMOS and PMOS transistor pairs to generate rich harmonics and tailor the ISF. This design approach minimizes the conversion of transistor noise into phase noise (PN) in both the $1/f^3$ and $1/f^2$ regions. While the use of dual transistor pairs results in doubled DC power consumption due to the higher supply voltage, it achieves a 3-dB reduction in phase noise. To prevent mode ambiguity and latch-up, a thin metal resistor is integrated within the single-turn multi-tap inductor. The article also derives the equations needed to determine the upper bound of this resistor's value. In another work [30] (75 to 83 GHz, 130 nm SiGe BiCMOS), the Colpitts architecture leverages $g_m$ boosting and an inductive tail. This can be achieved by combining the power efficiency and start-up properties of the cross-coupled and the impulse sensitivity of the Colpitts topology, thus increasing the small-signal loop gain of the proposed structure. Moreover, it has a low-gain secondary frequency control method varying the bulk-drain parasitic capacitors of the cross-coupled pair, and it uses an RF choke for biasing varactors, which reduces the AM-PM PN contribution of the noise source coming from the biasing resistor. [31] (61 GHz, 130 nm SiGe BiCMOS) uses the Colpitts topology with a second tunable varactor at the base of the transistor pair, while there is no additional current injection into the emitter of the main stage for linearization.

Since phased arrays are commonly utilized for enhancing link budget and directing beams in mm-wave applications, multi-phase approaches have been shifted toward varactor-less structures with interpolating phase tuning. A multi-phase LC oscillator, [32] (67.8 to 81.4 GHz, 65 nm CMOS), employs rotated phase tuning and uses transistor pairs as implicit phase shifters. Note that both transistors cannot approach $f_{max}$ simultaneously. Phase interpolation happens by changing the phase of the flowing current into the LC tank. Optimizing the number of stages and device sizes would result in larger phase differences, thus increasing the frequency TR. Additionally, a flexible phase shift can avoid oscillation mode ambiguity while improving PN and TR. [33] (27.57 to 33.07 GHz, 65 nm CMOS) proposes a wide range of phase shifts for different frequencies in a rotated phase tuning approach by using a common-source common-gate transistor, thus realizing negative resistance units and phase shifters for a varactor-less QVCO. [34] (26.2 to 30 GHz, 22 nm FDSOI) proposed an improved version for rotatory traveling-wave oscillators that suffer from flicker noise up-conversion while gaining multi-phases and benefiting from low PN. TL dispersion causes higher-order harmonics to travel faster than the fundamental, and adding tuning capacitors on the distributed stubs slows down the travel speed of higher-order harmonics. This intentional phase difference cancels the phase shift due to phase dispersion, thus avoiding flicker noise upconversion. Another approach in [35] (12.4 to 13.56 GHz, 65 nm CMOS) is based on superharmonic coupling, maintaining phase accuracy in eight-phase generation instead of conventional quadrature-phase generation. The design leverages a class-C VCO with loop-based superharmonic coupling, which simultaneously offers low PN, high phase accuracy, and high energy efficiency. It adds a series inductor in the control section, allowing loop coupling to function as a ring oscillator. This setup forces the second harmonic CM nodes of four oscillator cores to lock orthogonally, producing eight phases of the fundamental frequency suitable for multi-phase clock generation, while preserving

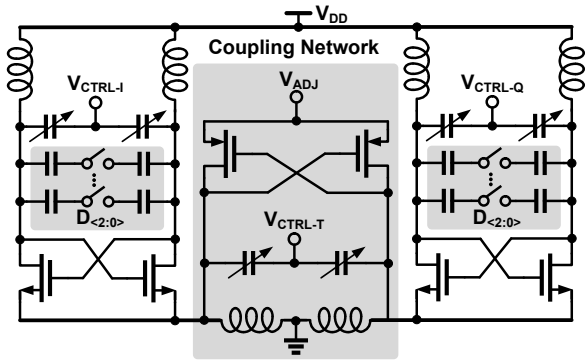

**FIGURE 4.** Superharmonic coupling scheme that allows quadrature error adjustment with negligible PN penalty by modifying the common tail impedance [36].

ISF and maintaining excellent PN performance.

In [36], a new mm-wave superharmonic-coupled QVCO topology is introduced in which coupling is performed at the second-harmonic CM nodes rather than at the fundamental nodes. This approach avoids direct loading of the LC tank, thereby preserving the TR, while also benefiting from the tail filtering effect to potentially improve phase noise as shown in Fig. 4. The authors developed a "tail impedance theory" showing that by engineering the tail impedance $Z_{Tail}(\theta)$ to peak at a desired phase difference, an arbitrary steady-state phase relationship, including perfect quadrature, can be achieved. In their implementation, a PMOS cross-coupled pair with a small tail inductor resonates at $2\omega$ to maximize $Z_{Tail}$ at $90°$, boosting the second harmonic at the common-source node. This enhanced second harmonic strongly modulates the effective $g_m$, impacting the loop gain and start-up conditions. Importantly, varying the coupling strength $g_{mp}$ allows post-fabrication tuning of the quadrature accuracy with almost no PN penalty, thanks to a noise-balancing effect between the NMOS and PMOS devices.

In [37] (10 GHz, 65 nm CMOS), a $g_m$ boosted VCO consisting of a center-tap XFMR and a stacked transistor core is included. Although the oscillation frequency is below the mm-wave range, the concept can be greatly extended to higher frequencies, and negative $g_m$ boosting improves PN through ISF manipulation while enhancing the start-up condition. The required start-up current is reduced in proportion to the $g_m$ boosting ratio, resulting in lower power consumption without PN degradation. Moreover, aligning the minimum noise modulation function with the maximum ISF value decreases the effective ISF, thus improving PN.

Rotary traveling-wave oscillator (RTWO) is a circuit architecture that produces multi-phase clocks within a closed-loop TL [38]. An RTWO often includes distributed inverters that serve as both amplifiers and latches to maintain the oscillation and ensure frequency locking [38]. In [39] (28.3 to 32.9 GHz, 65 nm CMOS), an LC-based segmented elementary circuit model is presented for RTWOs, and an extrinsic Q factor expression for the TLs is derived. Additionally, an extrinsic-Q-enhanced TL featuring a hollow ground plane is utilized in a sample oscillator, achieving very good performance compared to state-of-the-art multi-phase oscillators. In [40] (15.22 to 18.23 GHz, 28 nm CMOS), the mechanism of flicker noise upconversion in inverse-class-F oscillators considering the unequal parasitic capacitances of NMOS and PMOS transistors is analyzed. Additionally, a balanced dual-core inverse-class-F with reduced flicker noise upconversion and improved PN is implemented based on the analysis. In a novel method for constructing oscillators without using tuned resonators, [41] (29 to 36 GHz, custom EPI-layer design) suggested employing resonant tunneling diodes and their associated negative differential resistance in the I-V characteristic to create high-frequency oscillations. The proposed circuit can attain considerably higher DC-to-RF efficiency compared to traditional oscillators.

In [42] (19.64 GHz, 65 nm CMOS), a K-band VCO is presented that employs a phase shift between the gate and drain terminals. This results in a symmetry in the effective ISF and reduces the flicker noise corner. It is paramount when the synthesizer loop bandwidth is comparable to the $1/f^3$ PN corner. Two auxiliary LC tanks shift the relative phase of the fundamental tone between the gate and drain terminals and attenuate the second harmonic of the output waveform. In other words, it suppresses PN caused by flicker noise up-conversion without compromising the desired $1/f^2$ PN. It is noteworthy to mention that the loop from drain to gate provides a passive gain at $f_0$ due to the auxiliary tanks, thereby suppressing the thermal noise.

### C. UTILIZING PASSIVE RESONATOR CIRCUITS

Efforts are made to increase the coupling factor of the tank for improved PN. In [43] (tri-band at 18.4, 23.8, or 35.1 GHz, TR 17%, 28 nm CMOS), triple coil windings and a set of control switches are utilized to implement a tri-band VCO. Two metal layers are connected with many vias, as necessary to minimize Q reduction. The length of windings and overlaps is controlled to achieve the desired mutual couplings in a compact area. In [44] (37.3 to 38.6 GHz, or 38.3 to 40.7 GHz, 28 nm FDSOI), two pairs of varactors are enabled and disabled by a set of switches. Here the concept is that in either the upper or lower frequency band, the disabled varactors function as fixed capacitors to shift the operating frequency into the desired range, while the enabled varactors offer sufficient TR. In [45] (54.8 to 71.5 GHz, 22 nm FDSOI), a 10-bit current-steering DAC adjusts the coupling between the I and Q cores of a quadrature digitally-controlled oscillator (DCO) by varying the bias currents, thereby altering the oscillation frequency. This approach achieves a wide TR of 11 GHz in the designed ADPLL for frequency-modulated continuous-wave (FMCW) radar applications.

To achieve extremely low PN levels, developing an oscillator based on a series resonance tank has been proposed in the literature [15], [46]–[48]. In a notable work [47], the key insight is that at resonance, the series-tank resistance $R_S$ is $Q^2$ times smaller than the equivalent parallel resistance $R_P$ which allows the same tank voltage swing to dissipate $Q^2$ more power in the resonator, theoretically improving PN by $10\log(Q^2)$ dB over a parallel tank. To properly drive the series tank, a custom negative-resistance cell is devised as depicted in Fig. 5: a cross-coupled BJT pair ($Q_1$ and $Q_2$) acting as a differential emitter follower connected to the series tank, combined with diode clamps ($D_N$ and $D_P$) to enforce large-signal voltage saturation at the collectors. Additional measures minimize noise, including inductor-based biasing, $L_T$, at emitters to avoid resistor thermal noise, careful parasitic resonance tuning, and voltage distribution across multiple MOM capacitors and varactors $C_S$, to prevent reliability issues under the high voltages in the series-tank. The prototype achieves $-138\,\text{dBc/Hz}$ PN at 1 MHz offset and $-190\,\text{dBc/Hz}$ FoM, rivaling III–V technology performance while consuming modest power. It is worth noting that designs based on series resonators may draw substantial supply current due to the low tank impedance at resonance, particularly when a high-Q resonator is used to achieve lower PN. However, this potential drawback is often offset by the exceptionally low PN attainable with such architectures, leading to an excellent FoM despite the higher power consumption. Additionally, the larger voltage swings across the inductor and capacitor must be carefully managed (e.g., through capacitive voltage division) to avoid excessive stress on the junctions of the active devices.

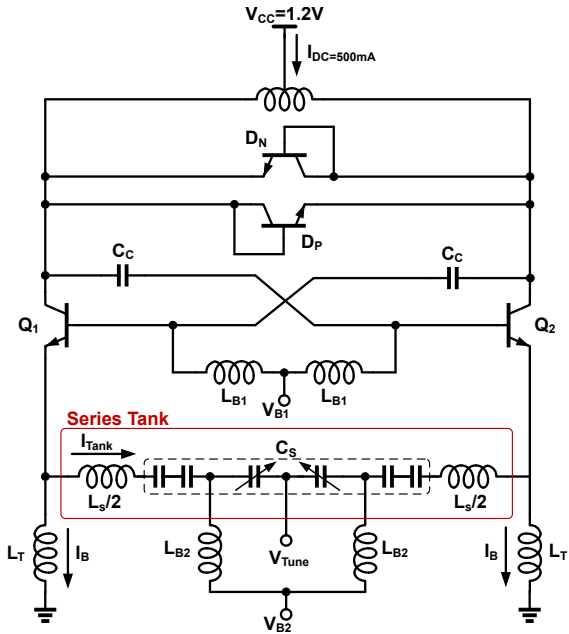

**FIGURE 5.** A series-resonance BiCMOS VCO with extremely low PN and high FoM [47].

A series resonance VCO may be utilized for quadrature signal generation. [48] (10.3 to 11.1 GHz, 65 nm CMOS) proposes a series resonance VCO featuring quadrature-phase outputs while occupying a small chip area. It comprises a folded structure based on an XFMR. However, achieving a wide TR with this method is questionable. An eight-core series-resonance VCO with a scalable ring-coupling scheme is reported in [49] (11 to 12.1 GHz, 65 nm CMOS) that uses 16 coils and achieves one of the lowest reported PN in a CMOS implementation. The problem of frequency mismatch in a multi-core approach is partially circumvented by a compensation mechanism at the cost of design complexity. However, this approach has limited scalability due to the need for high-voltage-tolerant devices because of the high voltages occurring at the internal nodes in series resonance VCOs.

Folded series tank topology is a method to reduce chip area by folding the XFMR trajectory inward, but it results in differently shaped XFMRs at corners and edges, causing tank mismatches between cores. The proposed mesh topology in [12] reduces chip area even further, to about half of the folded series topology, by efficiently using the inner grid edges for the XFMRs. This also reduces the ohmic drop due to shorter power/ground paths. Additionally, the mesh topology uses uniformly shaped XFMRs, minimizing potential tank mismatches. In [50] (7.1 to 16.8 GHz, 22 nm FinFET), a triple-mode VCO is presented to achieve an octave TR. Here, three distinct resonant frequencies are achieved through constructive and destructive magnetic coupling, coil-current cancellation, inductor shortcutting in a high-Q compact 4-port tapped inductor, and capacitive mode-switching. In [51] (25 to 28 GHz, 7 nm CMOS), two weakly coupled inductors are combined to create a 4-port inductor, enabling higher bias current and enhancing the VCO's FoM. A bowtie-shaped structure is implemented to minimize on-chip magnetic coupling. Additionally, a low-Q DC path is introduced to suppress CM oscillations and ensure high-Q DM operation.

In [52] (11.5 to 14.3 GHz, 65 nm CMOS), the head and tail inductors are coupled to cancel the noise injected from $V_{DD}$ and GND nodes due to their different polarity. In this design, the head inductor is also used as the differential mode resonant inductor to save area. Additionally, a CM-noise isolation XFMR is utilized to prevent the noise of the $V_{DD}$ from being injected into the gate of the active core transistors. In [53] (22.4 to 26.8 GHz, 65 nm CMOS) to address the trade-off between frequency-mismatch-induced PN penalty and the Q of the inductor (or XFMR), a dual-path-synchronized quad-core oscillator using a circular XFMR is proposed. This design introduces two coupling paths that simultaneously synchronize both adjacent and non-adjacent cores, effectively suppressing the PN penalty while maintaining the high-Q advantage of the circular XFMR.

In [54] (22.4 to 31.2 GHz, 45 nm SOI CMOS), the impact of switched-capacitor bank feedline on the performance of the mm-wave oscillators is investigated. Additionally, a double-folded feedline structure is proposed to linearize the coarse tuning characteristics and enhance the TR of the prototype VCO by reducing the parasitic capacitances. In [55] (21.1 to 25.5 GHz, 65 nm CMOS), a triple-coil XFMR-coupled quadrature VCO (QVCO) is presented in which the quadrature signal is provided without active-coupling transistors. Oscillation state analysis is performed to avoid mode ambiguity. The closed-form expressions for tank current and Q are derived, leading to design guidelines to improve the PN performance by utilizing large source inductances and strong coupling factors between the cores.

There have been efforts to replace the lumped inductor of the conventional LC resonator with other alternatives in mm-wave oscillators. In [56] (77.1 GHz, TR 4.79%, 55 nm CMOS), a VCO design using a slow-wave coplanar stripline-based distributed inductor is presented and compared to a conventional LC resonator-based VCO in the same process. The study concludes that with the distributed inductor, it is possible to achieve similar performance levels as the conventional LC resonator, with a slightly better TR and lower power consumption due to the relatively higher Q factor. In another work [57] (25 GHz, TR 5.4%, 180 nm CMOS) resonators based on *defected ground structures* are utilized to introduce transmission poles besides the poles of the conventional parallel resonators, thus improving the skirt characteristics of the S-parameters and leading to increased Q factor and enhanced PN. Alternatively, in [58] (76.09 to 78.6 GHz, 55 nm CMOS), a standing-wave oscillator (SWO) is proposed which relies on the distribution of varactors along an asymmetrical slow-wave coplanar stripline, which enables the improvement of the Q factor of the tunable resonator, resulting in superior performance in terms of PN and DC-to-RF efficiency at mm-wave frequencies. Another unconventional architecture utilizing a transmission-type Yttrium Iron Garnet (YIG) resonator is presented in [59] (19.1 to 41.4 GHz, and 32 to 48.2 GHz, YIG on SiGe). In this design, the resonator is realized by a YIG sphere confined between two orthogonal crossing bond wires on top of a SiGe chip. A circuit model for the resonator is derived, and the measurement shows excellent PN performance and FoM of 235 to 248 dB for two different implemented circuits. In [60] (27.9 to 28.8 GHz, 90 nm SiGe BiCMOS), the options in the design of XFMRs to replace the LC tank are studied, and a comprehensive model for the XFMR, including parasitics, is formulated. Additionally, a VCO utilizing a stacked XFMR and custom-made MOM capacitors is implemented and benchmarked.

In [61] (25.2 to 29.4 GHz, 65 nm CMOS), a multi-resonant *Resistor-Inductor-Capacitor-Mutual inductance (RLCM)* tank is employed within a single-branch VCO, allowing for reuse of the bias current to achieve low power consumption (i.e., 3.3 mW from a 1.1 V supply). Additionally, $RLCM$ is utilized to generate two high-$Q$ DM resonances at the fundamental and second harmonic oscillation frequencies to reshape the ISF and improve PN performance.

To further increase the TR, [62] (82 to 107.6 GHz, 65 nm CMOS) presented a DCO using a split XFMR for magnetic tuning. It also splits the XFMRs into two parallel ones to minimize the mutual coupling between the second coils. Similarly, [63] (57 to 90.1 GHz and 95.7 to 110.5 GHz, 65 nm CMOS) uses magnetic tuning and minimizes PN degradation, while combining XFMR-based magnetic tuning and varactor-based capacitive tuning increases the VCO's bandwidth. A highly linear and ultra-wideband VCO is presented in [64] (75 to 86 GHz, 65 nm LP CMOS) where two secondary coils are coupled to the varactor, and the triple-coupled XFMR shares the center tap.

A fully decoupled LC tank can enhance oscillation amplitude while maintaining low PN [65] (31.8 to 39.6 GHz, 130 nm SiGe BiCMOS). Unlike conventional LC-tank VCOs, this design connects the tank to the gate of the cross-coupled pair while keeping it electrically decoupled from the gates. By employing a decoupling technique, the design effectively reduces PN. The use of a capacitive XFMR isolates the LC tank from the active devices, allowing for higher oscillation amplitudes that are no longer constrained by the transistor breakdown voltage. This decoupling also minimizes noise injection from the active devices into the tank. Additionally, it increases the tank current and lowers the tank's equivalent resistance, leading to enhanced signal power. A fully differential VCO is presented in [66] (24.5 to 28.3 GHz, 65 nm CMOS) where a multi-LC tank utilizes two separate single-turn inductors and a 1D tuning capacitor. This capacitive-coupling VCO leverages wideband $1/f^3$ PN corner reduction since ISF does not shift across the TR. It provides high-Q high-impedance resonances at the fundamental and second harmonic and eliminates the CM resonance by a series gate resistor.

At the cost of larger size constraints for XFMRs, [67] (22.5 to 27.7 GHz, 28 nm CMOS) uses XFMR feedback between the drain and source of the cross-coupled pair with a tunable source bridge capacitor, which provides low PN and wide TR for a mm-wave DCO. Proper magnetic coupling increases the resonator Q, saving power for lower PN. Also, the source-bridged capacitor adjusts the phase difference between fundamental and second harmonics and suppresses flicker noise up-conversion, resulting in a lower flicker noise corner for the PN profile and a more symmetrical oscillation waveform.

### D. UTILIZING ADVANCED PASSIVE COUPLING NETWORKS

Seamless tunability is offered in a wide operating range LC quad phase VCO presented in [68] (14.3 GHz, 65 nm CMOS). It consists of two LC tanks with a series coupling to have the desired phase sequence and a parallel coupling for a wide TR. Moreover, its seamless tunability is achieved by changing the polarity of the parallel coupling network which does not require hard switching. Regardless of tunability, $g_m$ boosting can improve PN. Even changing physical layer structures could lead to improving an oscillator's efficiency. [69] (17.5 GHz, 180 nm CMOS) optimizes the ratio of $C/L$ to affect the load Q factor and eventually the PN. This enhancement happens by creating a defective ground structure (DGS) for the resonator. In other words, an H-shaped DGS on the thickest metal layer with small inductances enhances the Q-factor, a benefit not found in a conventional LC resonator.

Quadrature oscillators are commonly preferred for generating I/Q signals. However, conventional approaches such as hybrid couplers and polyphase RC filters are inherently narrowband, resulting in substantial phase errors in wideband applications. Although adding more stages, either through cascaded couplers or multi-section RC filters, can help minimize these errors, such methods are rarely adopted due to their high insertion loss, which significantly impacts power efficiency at mm-wave frequencies. To overcome the trade-off between low PN and small quadrature errors, observed in traditional quadrature VCOs, the work in [70] (24 to 29.2 GHz, 28 nm CMOS) proposes a coupled PLL-based CMOS quadrature VCO. This allows the PN to be improved by 3 dB within the PLL bandwidth. In the coupled PLL-based QVCO, each VCO acts as a reference for the other one, and quadrature accuracy is controlled by the control path parameters employing a mixer-based phase detector. Each VCO employs a class-B differential topology with a PMOS pair. To mitigate flicker noise up-conversion, the CM tank impedance has a resonance at the second harmonic frequency. This is achieved by coupling an additional inductor at the center of the tank inductor to the tail inductor, which increases the total inductance of the path for the second harmonic. To achieve quadrature phases at lower frequencies, [71] (4.9 GHz, 40 nm CMOS) proposes a quadrature VCO with a coupling based on a tail-switch network. It does not have an explicit parallel or series coupling network and avoids bimodal oscillation and noise due to coupling paths while providing accurate I/Q phases. To achieve a high FOM and a continuous TR, n-port resonators can be utilized in an oscillator. [72] (4 to 5 GHz, 22 nm FDSOI) proposes a VCO based on a 4-port XFMR resonator. This resonator is employed to provide two DM and two CM harmonic impedances, resulting in lower PN, voltage supply, and sensitivity to CM tuning. Not only does the class-$F_{23}$ tank provide a higher CM impedance at the second harmonic for reduced PN, but its source feedback paths also leverage gain boosting and $3^{rd}$ harmonic shaping for increased efficiency and reduced PN. Keeping an XFMR between drains and gates of the transistor pair provides a class-C XFMR coupled VCO [73] (18.14 to 21.23 GHz, 22 nm FDSOI). To make a compact design suitable for PLLs, the XFMR is implemented as overlapping single-turn coils in the top two thick metal layers, providing a higher quality factor with less eddy current.

In the case of a fixed supply, lowering inductance increases the tank power and thus improves PN, although the minimum size of the inductance is limited due to the Q drop. In [74] (25.1 to 28.3 GHz, 7 nm CMOS), a mm-wave VCO is presented in which lightly coupled inductors are employed. This leads to lowering the total inductance, but at the same time, it enables multiple resonant modes. A twisted four-port geometry is chosen to force the current through and select one mode unambiguously. Furthermore, two thin traces bypass the inductor, and the intentionally low-Q made DC path avoids latching the differential pairs on both sides of the inductor. Ring oscillators can be low-power and area-efficient when stringent power and area budgets matter. In [75] (23.9 to 26.9 GHz, 65 nm CMOS), a two-tank transformer (XFMR)-feedback topology is introduced to boost oscillation swing under low supply voltage. Additionally, flicker noise up-conversion is mitigated by setting the drain impedance of the cross-coupled transistors to a real value at the second harmonic of the oscillation. Configuring the equivalent source impedance of the cross-coupled transistors to be a real negative value enhances the loaded Q factor and reduces the noise contribution from the VCO's transconductor. A slightly modified version of [75] in a newer process is presented in [76] (22.2 to 29.1 GHz, 22 nm FDSOI CMOS). The design achieved better PN performance mainly due to the modified layout of the resonators along with a more advanced process technology. In [77] (20 to 23.8 GHz, 65 nm CMOS), an 8-phase oscillator featuring magnetic and dual-injection coupling is presented to relax the trade-off between the PN and phase error while improving the area efficiency. In this design, quad-core oscillators are coupled in a ring while having magnetic coupling through a symmetrical windmill-shaped XFMR. Dual tail injection using adjacent multiphase outputs is utilized to reduce the PN while enhancing the phase error between various outputs.

In [78] (24 GHz, 16 nm FinFET), a quad-based multi-ring oscillator, which includes embedded phase interpolators and voltage follower-based cross-coupling loops, is proposed. To push the DC-to-RF efficiency closer to the efficiency limit of the transistors, [79] (82.2 to 89.3 GHz, 65 nm CMOS) proposes a single-device-based VCO leveraging a device-centric method to achieve a high-efficiency oscillator. The structure consists of a class E power amplifier and an active feedback network that provides maximum swing at the gate of the core device, increasing the power added efficiency (PAE). It also uses interdigitated capacitors with a higher intrinsic self-resonance frequency, thereby improving the quality factor.

To enable the use of both the fundamental and second harmonic frequencies, [80] (12.3 to 15.2 GHz, 65 nm CMOS) introduces a low PN VCO featuring CM resonance expansion and an intrinsic differential second harmonic output, utilizing a single three-coil XFMR. By incorporating a third internal coil into the conventional XFMR-based class-F resonator, the CM resonance can be actively tuned. This approach eliminates the need for manual tuning to align the second harmonic resonance, resulting in a wider CM resonance bandwidth. Moreover, the design achieves simultaneous wideband differential resonance at the second harmonic, removing the need for a frequency doubler and thereby reducing power consumption and chip area. The proposed stacked 8-shaped inductor, functioning as the third coil, is integrated within the existing XFMR structure, avoiding any additional chip area overhead.

At lower mm-wave frequencies, a cross-coupled push-pull with source degeneration was used in [82] (16.3 to 19.7 GHz, 28 nm FD-SOI) to reduce flicker noise by controlling current flow in the tank. Another work [83] (23 to 29.9 GHz, 40 nm CMOS) suggested using coupled inductors in a circular geometry offering no mode ambiguity. Using coupled inductors, [84] (42.9 to 50.6 GHz, 65 nm CMOS) proposes a single-center-tapped switched inductor in which a quad-core coupled VCO leverages mode switching for achieving a wide TR and low PN. Similarly, [85] (49.2 to 56.6 GHz, 65 nm

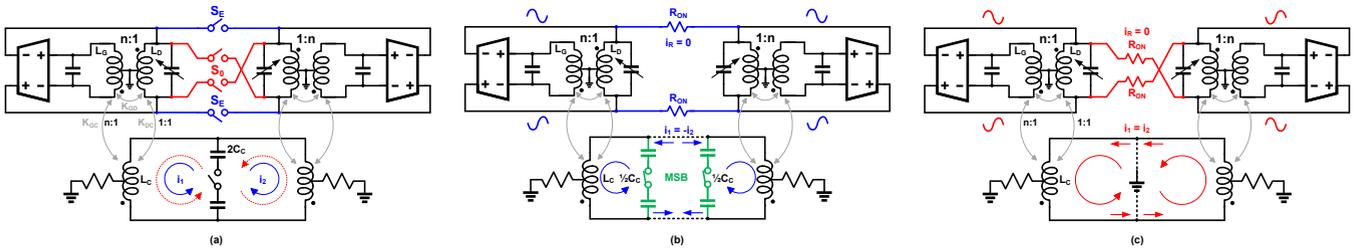

**FIGURE 6.** A mode-switching oscillator with two XFMR-based VCOs. This class-C coupled VCO utilizes mode switching and incorporates XFMR switching to achieve an octave TR, employing a switch-free tertiary magnetic coupling loop [81]. (a) The proposed VCO with its mode-switching network, (b) the VCO in the even-mode operation, (c) the VCO in the odd-mode operation.

CMOS) uses digitally-controlled inductors for a V-band mm-wave VCO that includes two switched secondary coils. It provides a wide inductor-based TR, and minimizing Q degradation in switched inductors enhances PN performance. A tri-band VCO is presented in [86] (33 to 41.13 GHz, 28 nm CMOS) that utilizes a three-mode reconfigurable inductor. It leverages magnetic tuning and adjusting the diameter of each winding, and its loop provides the equivalent inductance for each mode independently. The reconfigurable inductor includes primary and secondary windings with a center loop connected to each winding center tap.

Multiport coupled inductors offer interference suppression in the mm-wave LC VCO [87] (24 to 30 GHz, 7 nm FinFET). The design utilizes two single-turn inductors to form a coupled 4-port 8-shaped structure with enhanced EM behavior. Furthermore, it has two modes of oscillation from a dual-core oscillator with two LC tanks. Adding resistors between two closely spaced terminals of an inductor suppresses unwanted modes, thereby avoiding mode ambiguity. Applying inter-core shaping to multi-core oscillators provides robust CM resonance and cancels CM destructive couplings as reported in [88] (28 GHz, 40 nm CMOS). To ensure a high Q for the capacitor bank, differential capacitors are presented in both DM and CM. Its ability to cancel Cm destructive coupling increases the Q in both modes and enhances the second harmonic shaping. The inter-core shaping has low sensitivity to the parasitics in the signal's return path and to the mismatches between the cores, which enables a massive-core extension in designs with a larger number of cores.

### E. EXTENDED TUNING RANGE AND MULTI-MODE OSCILLATORS

In order to increase the TR of the mm-wave oscillators [89] (54.1 to 70.4 GHz, 65 nm CMOS) proposed a VCO having two switchable cores with overlapping frequency bands. When the first core is activated to produce the low-frequency band (LFB) signal and the second core is off, the inductors of the second core are repurposed to create additional buffers that pass the LFB signal to the output buffers. When the second core is turned on, it generates high-frequency band signals that are directly fed to the output buffers. Delivering the outputs of both VCO cores through the same terminals without using external active/passive combiners or coupled inductors enhances PN performance, increases output power, and reduces chip size. However, in this design, the untuned internal combiner reduces the power level of the LFB. In [90] (65 to 81 GHz, 65 nm CMOS), a wide TR is achieved through an XFMR-based variable inductor synthesis circuit. Here, a high-$Q$ capacitor in series with an almost lossless switch structure as the load of an XFMR is realized to build a dual-mode VCO which is able to achieve 22.8% TR around 73 GHz center frequency. In another attempt to increase the TR, in [91] (8.3 to 14.26 GHz, 65 nm CMOS), a boosted active-capacitor architecture using LC-tuned active-impedance-conversion is utilized as a tunable capacitance multiplier (tuning ratio of 7.8) and negative transconductance. In this work, dual supply voltages and a current reuse technique are used to reduce the power consumption. In [92] (10.8 to 19.3 GHz, 10 nm FinFET), a single XFMR-based LC oscillator with adjustable magnetic coupling is employed within an all-digital phase-locked-loop (ADPLL) to achieve a TR close to an octave. This is accomplished using the superior switches available in modern processes and the high turn-on/off capacitance ratio of the switched capacitors within the LC tank. In [93] (20.7 to 31.8 GHz, 28 nm CMOS), a dual-mode, waveform-shaping quadrature VCO is used in a PLL to obtain a wide TR and low PN. In the proposed circuit, mode switches, capacitors, and coupled inductors are utilized to couple each oscillator pair operating in-phase (even-mode) or out-of-phase (odd mode), generating the dual-mode operation. As another example, in [94] (55 GHz, TR 18%, 65 nm CMOS), a triple-coil structure is utilized to widen the TR by switching the coupled coils to the primary coil of the tank, varying the effective inductance of the resonator. In [95] (13.7 to 41.5 GHz, 65 nm CMOS), four oscillator cores are coupled together using two (main and aux) 8-shaped coupling tanks to form a quad-mode oscillator with a 101% TR. Various modes are selected using three sets of low-loss switches.

In [96] (8 to 11.15 GHz, 65 nm CMOS), a novel method is introduced to reduce both flicker and thermal PN contributions in LC oscillators over a wide TR in a relatively lower frequency band. This technique utilizes dual CM resonance, combining implicit and explicit CM resonances applied to the LC tank and tail node of the oscillator, respectively. This approach provides high resistive impedance across the second harmonic band without requiring dedicated tuning. A Colpitts-based differential LC tank can be used in a class-B oscillator with a single varactor and an inductor to make a wide-tuning VCO [97] (31.8 to 40.8 GHz, 40 nm CMOS). Although the TR is limited to $C_{max}/C_{min}$ of the varactors, the push-pull class-B amplifier, when directly coupled to the VCO with capacitors, reduces dependency and introduces additional negative resistors, enhancing robustness against severe variations in PVT such as near triode-biased class-C VCOs.

Apart from harmonic coupling, coupling through mode switching was presented in [81] (8 to 17 GHz, 22 nm FD-SOI), where two XFMR-based VCOs were coupled through a network of mode switches as shown in Fig. 6(a). It not only enables mode switching from the mode selection and tuning network without added parasitics, but also utilizes XFMR switching without switch loss or Q degradation. This results in an increased frequency TR without suffering from a considerable PN penalty. In the scheme of dual VCOs, DC coupling to the inductive-degenerated positive feedback buffer provides additional negative resistance to the tank, and it, even so, reduces the risk of bimodal oscillation. In the even mode as depicted in Fig. 6(b), the cores operate in-phase and therefore, the induced currents in $L_c$ are in-phase. In this mode, turning MSB on extends the low-band TR through magnetic coupling to the main XFMR. In the odd mode [see Fig. 6(c)], the cores operate in anti-phase. The induced current in the $L_c$ loop circulates with minimum loss, resulting in a smaller inductance for the main tank. Apart from isolating the gates of the cross-coupled pair and coupling them to the tank through an XFMR, explicit capacitors between the drain of each transistor and the common source are connected to improve the noise performance in this class-C XFMR-coupled VCO. A dual-mode voltage waveform-shaping oscillator is presented in [98] (20.7 to 31.8 GHz, 28 nm CMOS) in which there are two XFMR-based resonators coupled through two-mode switch circuits, generating both fundamental and third harmonic in two modes. The dual-mode switch circuits (without Q degradation) give square wave voltage outputs, thereby reducing ISF for lower PN and extending the TR. Utilizing mode switching, [99] (12.1 to 32 GHz, 65 nm CMOS) presents a wideband mode-switching quad-core VCO employing a compact multi-mode magnetically coupled LC network and providing four different resonant modes. This design manages the trade-off between the tuning range and the PN under a limited power budget. Furthermore, it leverages mode-switching circuitry to avoid multi-resonance oscillation and eliminate the mismatch between multiple VCOs.

The VCO in [100] (12.9 to 24 GHz, 65 nm CMOS) achieved a wide TR via efficient mode switching by using a multi-coil XFMR and an inductor network where the coils can be reconfigured through MOS switches to form either series-aiding or series-opposing magnetic coupling modes. In low-frequency mode, the inductors are connected to maximize total inductance, lowering the oscillation frequency while maintaining high Q. In high-frequency mode, the connection reduces total inductance and shifts parasitic reso-nances, allowing operation at nearly double the frequency without degrading tank quality. Because the switching only changes inductor connections (not adding lossy elements into the tank), the PN performance remains high in both modes. As shown in Fig. 7, the DM virtual ground nodes of the drain, gate, and source coils from two multi-core VCOs (e.g., core 1 and core 2) are separated and then interconnected with their counterparts in the active cores. After this reconfiguration, the two active cores share a common multi-coil XFMR tank, resulting in a 3 dB PN improvement through dual-core coupling. Similarly, [101] (25 to 35.9 GHz, 65 nm CMOS) uses coupled cores with mode switching to present a dual-layer, quad-core, dual-mode VCO featuring low PN, wide TR, and no issues with standard voltage reliability concerns. The multi-mode approach uses magnetic coupling between inductors seated in different layers that supports a higher VDD in an NMOS-only oscillator without reliability issues. In addition, the mode switches consist of in-layer inter-core connections and inter-layer inter-core transmission gate switches.

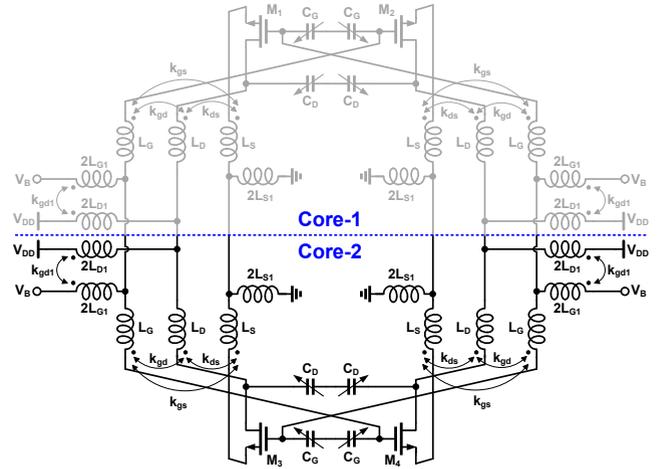

**FIGURE 7.** A dual-core VCO with mode switching utilizing a multi-coil XFMR for reduced PN and increased TR [100].

Utilizing tunable transmission lines (TLs) helps build oscillators with an ultra-wideband frequency range. [102] (3.1 to 51 GHz, 28 nm FDSOI CMOS) presents a compact, lumped/distributed LC-equivalent resonator covering more than four octaves in continuous frequency tuning. To achieve this, the cross-coupled LC VCO uses a tunable TL as the inductor for the VCO and avoids costly array-based large-area designs; however, one should not expect a very high Q from such a design.

Combining a multi-core approach with an XFMR-switching technique can provide mode design freedom. In [103] (20.65 to 40.55 GHz, 65 nm CMOS), the VCO features a dual-core design and four operational modes, incorporating a switched XFMR. It includes two extra modes that expand the octave TR without requiring additional cores. Moreover, the mode-independent switching cuts the off-capacitances of the switched XFMR. The 8-shaped inner coil splits the even modes into two, self-canceling the coupling effect in the even mode, and the outer coil splits the odd mode and cancels the coupling in the odd mode. It is important to note that the area of the proposed quad-core VCO is comparable to that of a dual-mode VCO due to the overlapping of its additional coils with the main coil. Leveraging an XFMR to couple the varactor to the LC core magnetically increases the TR by doubling the bias range as presented in [104] (20.8 to 28 GHz, 65 nm CMOS). This class-C VCO reduces the parasitic capacitance seen by the varactor and boosts the tank's Q due to impedance transformation, thereby improving both PN and frequency TR simultaneously. The TR, however, decreases partially due to the XFMR coupling.

## IV. DESIGN TECHNIQUES FOR MM-WAVE HARMONIC-TONE HARMONIC OSCILLATOR DESIGN

Another key category of oscillators used in the mm-wave band is harmonic-based harmonic oscillators that operate at the harmonics of the fundamental tone and are commonly referred to as harmonic oscillators. These oscillators offer a practical and efficient means of generating and manipulating signals within a frequency range that would otherwise be difficult and costly to achieve directly. As a result, harmonic oscillators play a crucial role in ensuring signal purity for systems operating at mm-wave frequencies. They provide control over a broad frequency range, enabling efficient signal generation and power amplification, while also maintaining strong performance in terms of PN and stability. Additionally, they offer versatility in modulation and signal processing. However, harmonic oscillators do come with some significant challenges, such as nonlinearity, harmonic distortion, temperature sensitivity, and so on. This section reviews recent solutions from the literature aimed at addressing these challenges to ensure efficient and reliable operation within the mm-wave frequency bands. Note that, similar to the previous section, some designs utilize two or more techniques, and it is discussed only in the subsection where it is most relevant.

### A. MULTI-CORE TECHNIQUES

A tightly coupled dual-core VCO with an implicit 4$^{th}$ harmonic extraction technique is reported in [105] (47.9 to 56.4 GHz, 65 nm CMOS) in which both fundamental and fourth harmonics are present at the output simultaneously. One direct application of this approach is the use of the fundamental tone for PLL locking. The overall jitter in the tightly coupled VCOs is enhanced due to an increase in the Q factor and a reduction in PN degradation caused by capacitor mismatch. The 4$^{th}$ harmonic extraction is based on passive components. In the CM operation, the power supply parasitic tap inductance and TLs create a step-up XFMR, resulting in a large passive gain for the 4$^{th}$ harmonic. Similarly, during DM operation, the coupling XFMR forms a dual-core resonator that concentrates DM energy, utilizing two fully symmetrical nested single inductors with a high coupling coefficient. As a result, the total capacitance experienced by the mutually coupled inductor remains constant, helping to minimize frequency drift and PN variations.

Combining a multi-core approach with switch-less mode switching overcomes PN degradation due to the switch's on-resistance. The VCO in [106] (7.1 to 15.7 GHz, 28 nm CMOS) presents a switch-less dual-core triple mode oscillator with magnetic and inductive tuning elements. To save power in triple modes, it only uses two negative $g_m$ cells. Moreover, to overcome the switch's on-resistance, the switching $g_m$ technique is used to select modes without using MOSFET switches, which are also naturally resistant to magnetic pulling because of their symmetrical footprint. To enable and disable $g_m$ cells, a tail transistor is used in each $g_m$ cell. Similarly, in [107] (57 to 69 GHz, 45 nm SOI CMOS), a dual-core and a quad-core, XFMR-coupled, second-harmonic VCOs are presented that achieved competitive PN performance at the mm-wave band.

Harmonic power manipulation, waveform shaping, and synthesis methods are utilized to implement a second harmonic VCO with improved FoM at mm-wave frequencies. At the 60-GHz band, a dual-core push-push Colpitts VCO presented in [108] (60 GHz, 130 nm SiGe BiCMOS) uses second harmonic extraction by mode separation. Signal extraction is used without loading the VCO core to avoid PN degradation. In DM, two cores are coupled magnetically at the fundamental frequency, while in the CM, the cores are coupled out capacitively at the second harmonic frequency. A complementary class-B topology is used in [109] (18.9 to 22.3 GHz, 55 nm BiCMOS) where two coupled resonators tied to each resonator tail increase the amplitude of oscillation relative to the supply voltage and reduce the flicker noise up-conversion from the tail current generator. Its second harmonic, occurring at twice the main resonator frequency, is extracted from the tail resonators by leveraging their higher quality factor, enabling intrinsic frequency multiplication within the tank. The resonator incorporates two transconductance-based tail filters along with a third inductor, which eliminates the frequency mismatch between the two tail resonators and provides a uniform tail impedance in a compact layout.

### B. WAVEFORM SHAPING TECHNIQUES

[110] (27.3 to 31.2 GHz, 28 nm CMOS) reports a low-flicker-noise class-$F_{23}$ oscillator leveraging implicit resonance in class-$F_{23}$ and explicit CM return path. It is mainly a class-F design that applies third-harmonic boosting and extraction techniques to maintain a high Q-factor for the tank. It uses proper terminations for harmonics as well as second harmonic resonance to limit flicker noise upconversion. For the second harmonic resonance, it embeds a decoupling cap inside the XFMR so that it makes an explicit CM current return path, and this current enters a resistive path, thereby reducing noise upconversion due to maintaining symmetry between rising and falling edges of the output waveforms. In a related study reported in [111] (28 GHz, 28 nm CMOS), the compromise of flicker noise upconversion due to asymmetry

in the oscillating waveform is addressed. The proposed solution involves flattening both the rising and falling edges of the voltage across the drain and source by introducing capacitive paths and employing inductive termination of the harmonic current, respectively. Similar to others, for $1/f^3$ PN reduction, narrowing the conduction angle resulting in a gate-drain phase shift is achieved by tuning the termination of the second harmonic current. [112] (57.5 to 67.2 GHz, 28 nm CMOS) presents a DCO leveraging reduced flicker noise up-conversion to the $1/f^3$ PN profile. It cancels out the fundamental tone, and the third harmonic is extracted and delivered to the output, where the XFMR-based dual resonance LC tank provides a large impedance to boost the 3$^{rd}$ harmonic. Due to phase modulation by the CM component of the flicker noise, adding a large impedance at DC in series with the $g_m$ pair, i.e., a current source, suppresses flicker noise up-conversion. The series branch between the supply (center tap of the XFMR) and the common-source node of the gm pair resonates at the second harmonic. The LC network between the drain and the source of the gm pair also resonates at the second harmonic in the CM operation. The former suppresses the direct up-conversion of the flicker noise, while the latter suppresses the indirect up-conversion of the flicker noise. In [113] (3.49 to 4.51 GHz, 65 nm CMOS), an inverse class-F VCO without a CM resonance circuit is presented. In this work, an XFMR-based dual-band LC resonator generates two intrinsic high-Q resonances at $f_{LO}$ and $2f_{LO}$, leading to low PN in the $1/f^2$ and $1/f^3$ regions while the XFMR helps reduce the effect of thermal noise of the used switches. The high-Q resonance is the reason behind the very high FoM at various offset frequencies, reaching a value of $196 \, \text{dBc/Hz}$.

In [114] (26 GHz, 28 nm CMOS), there is a 3$^{rd}$ harmonic extractor followed by a buffer from a class-F quadrature DCO whose XFMR couples the phase delay to avoid quadrature uncertainty. Exploiting the source-combining technique in a triple push VCO as depicted in Fig. 8 enhances the third harmonic [115] (52.3 to 67.3 GHz, 65 nm CMOS). Depicted in Fig. 8(a) is the single-ended structure of the proposed triple-push oscillator. The detailed implementation is shown in Fig. 8(b), where the 6-phase outputs are combined through sources of the $g_m$ devices, boosting the third harmonic current and inherently rejecting the first and second harmonic spurs. The design suppresses flicker noise up-conversion by adding a resistor to the center tap of the primary coil of the output XFMR. Additionally, the design addresses the common problems of off-resonance and bimodal oscillation in triple-push oscillators by coupling the auxiliary $g_m$ cells, ensuring that the voltage and current in the tank are in phase. In [116] (21.7 to 26.5 GHz, 28 nm LP CMOS), a quadrature 3$^{rd}$ harmonic oscillator based on a source-coupling Class-F architecture is presented which utilizes harmonic extractor circuits to amplify the third harmonic while suppressing the fundamental tone.

ISF manipulation can also be employed in harmonic oscillators. [117] (76 to 82 GHz, 65 nm CMOS) presents a super-harmonic VCO based on a class-E resonator employing harmonic-assisted ISF manipulation. Enhancing PN is accomplished by significantly increasing the second harmonic while reducing the thermal noise contribution from the fundamental signal. In [118] (75 to 86 GHz, 120 nm SiGe BiCMOS), a frequency quadrupler circuit is employed for mm-wave signal generation. The quadrupler circuit is fed with differential quadrature inputs that are provided either by a quadrature VCO or a tunable active polyphase filter.

### C. INNOVATIVE PASSIVE CIRCUITS

In [119] (54.9 to 63.5 GHz, 65 nm CMOS), a current reuse and current output VCO is integrated with a passive harmonic-current filter, designed to reject the first and second harmonics, and a 60 GHz trans-impedance amplifier (TIA). This combination achieves a large output power swing while maintaining low power consumption.

Coupling fundamental and second harmonic VCOs can generate a lower PN profile than a standalone second harmonic oscillator. [120] (23.9 to 29.4 GHz, 10 nm FinFET) reported a coupled frequency doubler for its proposed LC PLL. There are two oscillators at 14 GHz and 28 GHz coupled by an XFMR. This coupling provides extra PN reduction due to the superior PN of the former VCO compared to the latter. In practice, this architecture of stacking two oscillators reuses the current, thereby lowering power consumption. Moreover, employing the XFMR amplifies the output signal power by coupling the second harmonic tail current of the fundamental oscillator to the inductor current of the second harmonic oscillator. To achieve higher TR and maintain lower PN across the whole range, the self-mixing approach can be used [121]. In [122] (51.5 to 62 GHz, 65 nm CMOS), a self-mixing VCO topology to mix the fundamental frequency with the second harmonic is presented. Additionally, a magnetically coupled varactor to the oscillator core through an asymmetric XFMR is proven to limit the drop in the Q factor of the tank, as well as reduce the ratio of parasitic capacitance to total tank capacitance. An even larger Q can be achieved by operating the varactor in accumulation mode. Consequently, the TR and PN are improved compared to traditional tanks in LC VCOs.

The CM resonance expansion technique can also be applied to lower frequencies in the mm-wave band. [123] (25 to 30.7 GHz, 65 nm CMOS) proposes a VCO with simultaneous differential 2$^{nd}$ harmonic output while using only a single 3-coil XFMR. It adds an extra resonator to the main tank. This approach improves PN over a wide bandwidth at the cost of increasing the VCO footprint. Moreover, it employs a stacked 8-shape inductor and a low-k XFMR and generates differential 2$^{nd}$ harmonic output without requiring an extra frequency doubler. It is important to note that the third coil is coupled to the drain in the CM and remains unaffected in the DM. In [124] (75, 105 GHz, 40 nm CMOS), a dual-band VCO utilizing CM resonance expansion to suppress flicker noise upconversion without the need for harmonic tuning is presented. A harmonics-rich class-$F_{23}$ XFMR, which has DM resonance frequencies at $f_0$ and $3f_0$ and CM resonance frequency at $2f_0$, is used to generate two frequency bands in a compact layout featuring low PN. Here, the $f_0$ signal is coupled by an XFMR to a frequency doubler to form the second harmonic. The XFMR is reused to introduce additional CM resonance in the tank to broaden the CM range around $2f_0$, effectively reducing the $1/f$ noise upconversion. Additionally, $3f_0$ is buffered as an additional output of the dual-band VCO.

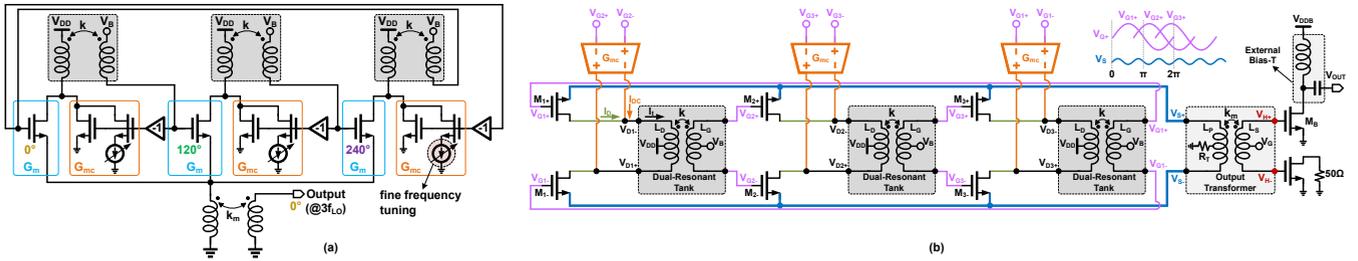

**FIGURE 8.** A triple-push VCO employing the source-combining technique to enhance third harmonic current. It uses coupled auxiliary $g_m$ cells to maintain the current and voltage in phase and avoid bimodal oscillations [115]. (a) The single-ended structure of the proposed triple-push oscillator, (b) the detailed implementation.

## V. DESIGN TECHNIQUES FOR SUB-THZ AND THZ FREQUENCY OSCILLATOR DESIGN

Generating oscillations above 100 GHz becomes increasingly challenging as the issues encountered in mm-wave oscillators remain relevant, along with additional physical and technological limitations of current technologies. These include device performance, propagation losses in interconnects, frequency drift due to temperature variations, and electromagnetic interference (EMI) emissions [125], [126].

Among the various techniques for harmonic-tone harmonic oscillators at mm-wave and sub-THz/THz frequencies, push-push and triple-push oscillators have garnered attention for generating the second and third harmonics, respectively. These two categories are fundamentally different in their operating principles and symmetry conditions. Push-push and triple-push oscillators utilize device nonlinearity and phase symmetry to generate high-order harmonics. In both cases, the oscillator cores operate at the fundamental frequency, and harmonic generation is enforced by symmetry rather than filtering. In a push-push oscillator, two cores with a 180° phase difference cancel the fundamental at the output node, while their second-harmonic components, in-phase due to the 360° phase shift, reinforce each other. This inversion symmetry enables efficient second-harmonic generation while suppressing the fundamental. On the other hand, a triple-push oscillator uses three cores spaced 120° apart, achieving three-fold rotational symmetry. This cancels the fundamental and second harmonic, as the 240° phase shift between second-harmonic components results in zero vector summation. The third harmonic, however, adds constructively because tripling the 120° phase shift yields 360°. Thus, push-push oscillators reinforce even harmonics, while triple-push oscillators reinforce harmonics whose order is a multiple of three. Note that, at these frequencies, maintaining symmetry is critical, as phase and amplitude mismatches, along with layout parasitics, can degrade harmonic purity. This makes triple-push oscillators, in particular, more sensitive to these issues due to stricter phase alignment requirements.

### A. GENERAL FUNDAMENTAL TONE OSCILLATORS

Although using the fundamental tone for mm-wave oscillation has been demonstrated in recent work, it is still not the predominant approach due to the difficulty in sustaining oscillation at mm-wave and THz frequencies. This is more challenging in the THz band, where the oscillation frequency is in the vicinity of $f_{max}$, even with advanced technology nodes. The limited $f_{max}$ pushes toward the use of higher harmonics, while the need for higher power generation encourages the use of power-combining techniques. However, this leads to degraded DC-to-RF efficiency and larger form factors [127]. [128] (150 and 245 GHz, 40 nm CMOS) proposes maximizing the power-area efficiency of a fundamental oscillator while preserving high DC-to-RF efficiency performance. It uses large-sized devices while minimizing effective parasitic capacitors. In each stage of this multi-stage oscillator, a phase shifter, serving as the passive network, is followed by a large transistor acting as the active network. Additionally, a pair of TLs is incorporated into the power-combining network to achieve optimal load matching. In [129] (292 to 318 GHz, 130 nm SiGe BiCMOS), an oscillator with push-push topology employing CM impedance enhancement is presented. The technique entails maximizing the CM impedance by creating a parallel CM resonance at the second harmonic of the fundamental oscillation frequency. This proposed method is straightforward to implement, yet it significantly enhances output power and efficiency with minimal impact on tuning bandwidth. The capacitive degeneration is used in [130] (270.3 to 279.3 GHz, 130 nm SiGe) to shape the output conductance of the cross-coupled pair. Additionally, it employs inductive feedback to induce large swings at the transistor gates, couples oscillators and combines output powers.

In [131] (93.4 to 104.8 GHz, 65 nm CMOS), a method is proposed to improve the PN of the conventional in-phase injection-locked QVCO. In the proposed circuit, two identical VCOs are used along with a coupling network consisting of a pair of XFMRs and four back-to-back diode-connected transistors. The coupling network guarantees accurate in-phase coupling through proper steering of the currents, leading to improved PN. In [132] (169.6 GHz, TR 21.7%, 28 nm CMOS) to relax the trade-off between PN and TR, a hybrid mode-switching architecture is proposed. In this circuit, inductive mode switching is added on top of the conventional capacitive mode switching. The advantage of the proposed inductive mode switching is that the mode switches are implemented outside of the inductor ring; hence, they will introduce loss and degrade the Q of the resonators, resulting in enhanced TR with negligible PN penalty.

In [133] (253 to 280 GHz, 40 nm CMOS), two techniques are introduced to enhance the oscillation frequency and TR. The capacitance-splitting feedback method boosts both the resonance frequency and the negative transconductance of the VCO by isolating the gate and drain parasitic capacitances and assigning them to separate nodes. Additionally, the tunable source-degeneration technique adjusts the effective capacitive loading of the VCO core, resulting in an extended TR. In [134] (240 GHz, TR 1.7%, 22 nm FDSOI), a $\pi$-type topology is employed in the fundamental-tone oscillator to decouple the biasing of the gate and drain nodes, providing flexibility in optimizing oscillation conditions through load-pull simulations. This approach results in a compact design with excellent FoM and high DC-to-RF efficiency.

Switch-less reconfigurable VCOs at frequencies above 100 GHz are proposed as a candidate for future 6G communications. [135] (102.33 GHz, 28 nm CMOS) proposes a triple-push and push-push dual-band VCO including a clover-shaped inductor suitable for the W band. To change its mode between triple-push and push-push operations switch-lessly, a clover-shaped inductor with a three-port connection is used that provides a single-frequency multiplied output port configuration. Achieving superior TR is provided by switch-less frequency band shift, i.e., turning on/off the cores independently while not loading switches in the RF path. Similar to prior works, [136] (213 GHz, 65 nm CMOS) proposes a capacitive XFMR-based approach for designing fundamental oscillators above $f_{max}$/2 of the active devices while taking full consideration for nonlinearity and co-simulating source- and load-pull for optimal source and load impedances. Apart from popular SiGe BiCMOS technology for +250-GHz oscillators, [137] (266 to 273 GHz, 28 nm CMOS) uses the CMOS technology to achieve this, and the proposed fundamental VCO uses a coupled inductor to realize a nested structure.

### B. COLPITTS-BASED FUNDAMENTAL TONE OSCILLATORS

Although the Colpitts topology experiences higher power loss, its simplicity and reliable operation motivate designers to tackle challenges in the sub-THz range. The work in [138] (106 GHz and 148 GHz, 130 nm SiGe HBT BiCMOS) presents an above-100-GHz oscillator optimized for oscillation frequencies close to maximum frequency and uses a well-known Colpitts topology. It couples identical oscillators for PN enhancement. In addition, it employs both resonant biasing and CM coupling schemes to improve PN. However, the loss of active devices limits not only the maximum frequency but also the PN significantly at those frequencies above 100 GHz. Utilizing a differential common-collector Colpitts, [139] (180 GHz, 130 nm SiGe BiCMOS) presents a super-regenerative oscillator leveraging efficient phase recovery. It also uses a differential common-base output buffer boosting the output power while providing sufficient reverse isolation. Nonetheless, its power consumption of 8.8 mW is considered high for the reported low output power with its 6.5% TR. Coupling common-drain Colpitt structures as a single-ended core in a loop is presented in [140] (148 GHz, 22 nm FDSOI). In this 4-core coupled-loop oscillator, the coupling signal is fed into the source terminal of the core transistor in each stage. To enhance oscillation within the coupling loop network, the phase compensation capacitor is carefully chosen to optimize performance. Each core's output power is AC-coupled to the next stage, while a grounded coplanar waveguide serves as the tank inductor.

Apart from the Colpitts structure for +100 GHz, an XFMR-based D-band LC VCO is presented in [141] (140 GHz, 28 nm CMOS), where inductors are inserted into the drain-to-gate connections to boost both the output power and the loop-gain while lacking linear tuning capability. Using TLs as tank inductors along with long varactors and the existing parasitic base-emitter capacitors, a D-band Colpitts VCO is presented in [142] (130 GHz, 130 nm BiCMOS). L-section matching networks are used to maximize swing without compromising the headroom and PN, while biasing the transistors at the maximum $f_T$ current density maximizes the oscillation condition. The TL at the emitter provides a noise-shaping function with the varactor, and the shunt capacitor with the bias resistor minimizes high-frequency noise injection by the resistor.

### C. UTILIZING FREQUENCY MULTIPLIERS

In [143] (312 GHz, TR 4.8%, 40 nm CMOS), an oscillator-doubler architecture is proposed in which the doubler obtains the optimum load impedance at the fundamental frequency to generate the second harmonic output without instability issues. The circuit achieves good DC-to-THz efficiency due to optimized voltage waveforms. The antennas are also fabricated on-chip to enable sub-THz signal radiation. In [144] (153.9 to 195.3 GHz, 65 nm CMOS), a fundamental tone VCO centered around 19 GHz is implemented, and its output is fed into an injection-locked frequency tripler to generate a signal centered at 57 GHz. A third harmonic enhancement and extraction technique is proposed to achieve the final sub-THz signal centered at 171 GHz. In [145] (90, 180, 360 GHz, 90 nm SiGe), two additional signal sources based on a Colpitts-Clapp VCO at 90 GHz and a TR of 26.6% are presented. The second source comprises a frequency doubler and a wideband PA to achieve a TR of 28.8% at 180 GHz, while the third design, including an additional doubler, can achieve a TR of 29.9% centered around 360 GHz. The third design includes an integrated wideband hybrid coupler to generate quadrature output signals. The authors claim their work is the highest frequency TR achieved for any published silicon-based, fully integrated signal source.

### D. HARMONIC EXTRACTION METHODS

In addition to frequency multiplication, harnessing power from the harmonics of the fundamental frequency offers a promising approach to achieve high-frequency oscillations without compromising output power or PN. Among these, third harmonic extraction has gained more attention despite being particularly challenging. This subsection reviews recent design strategies developed to address the difficulties commonly encountered by designers in this area. [146] (154 to 195 GHz, 65 nm CMOS) uses a cascaded LC oscillator with an injection-locked multiplier in which an XFMR-based dual tank resonator is employed for third harmonic extraction. To enhance the third harmonic amplitude, the oscillation waveform at the output was prevented from being flattened even when the active devices are in the triode region. Adding a third coil in the XFMR at the source of the cross-coupled pair can increase the tank resistance at the third harmonic. The voltage from the source may drop below ground, which raises the overdrive voltage and thereby enhances the effective transconductance. [147] (127.5 to 144 GHz, 16 nm FinFET) reports a harmonic oscillator employing a class-F type with their harmonic extraction in which buffers are used for both the fundamental and the third harmonic. Two main adaptations are proposed for the conventional structure. To maintain the start-up and oscillation conditions, two cross-coupled PMOS transistors are added on the primary side of the XFMR. Also, the size of NMOS transistors is reduced to lower the size of the capacitor on the secondary side. In addition, the TR is controlled by a switched leakage inductor, which removes the need for a large tuning capacitor. In [148] (210 GHz, TR 5%, 90 nm CMOS), a triple-push third-harmonic ring oscillator operating above the $f_{max}$ of the process technology is presented. The idea is to couple two uneven-sized LC ring oscillators in reverse direction to effectively generate the third harmonic and lower the PN. In [149] (584.8 to 614.3 GHz, 65 nm CMOS), several oscillator cores based on [150] are coupled in parallel, and the third-order harmonic is extracted to achieve higher output power at 0.6 THz. A wideband shared patch antenna is utilized to reduce the sensitivity to shifts in frequency. Measurements with a lens are conducted to demonstrate higher output power levels.

### E. UTILIZING SELF-OSCILLATING HARMONICS-BASED MIXERS

The work reported in [151] (240 GHz, 65 nm CMOS) employs a self-oscillating harmonic mixer to generate the second harmonic from a 120-GHz fundamental signal. The mixer consists of two self-sustained oscillators connected via central slotlines. Within each oscillator, the slotline supports only the odd quasi-TE mode, resulting in out-of-phase voltages across its two conductors. This configuration forces the two transistors to oscillate in differential mode at the fundamental frequency.

### F. COLPITTS-BASED HARMONIC TONE OSCILLATORS

Leveraging Colpitts's simplicity, second and third harmonic Colpitts' are often found in the recent literature [152]–[155]. Second harmonic power generation from devices in the triode region can be attractive, especially if it affords near 300 GHz. [152] (293 or 298 GHz, 65 nm CMOS) presents a harmonic oscillator with second harmonic power generation. For continuous TR, it adds switches at the source of the transistor in the Colpitts structure, which can short the source to the ground when the switch is on or act as source degeneration when the switch is off. In [153] (205 GHz, 65 nm CMOS), the optimum condition for efficient second harmonic signal generation based on polyharmonic distortion is to maximize nonlinearity by increasing the swings at the gate and drain terminals. In doing so, inductors are added between the gate of one side and the drain of the other. Also, the TL line is added with capacitors at the source of the cross-coupled pair to filter the second harmonic voltage at the source. Additionally, the TL line mitigates the effects of parasitic inductances introduced by the layout of the return current path. Similar to [153], for maintaining high DC to RF efficiency, [154] (245 GHz, 65 nm CMOS) uses a push-pull differential Colpitts structure to build an oscillator that uses common-drain for second harmonic generation. Even the Colpitts structure can be optimized for the third harmonic, as [155] (280 GHz, 65 nm CMOS) uses inductors in all terminals of active devices.

### G. COMBINING MULTIPLIERS WITH HARMONICS

Combining multiplication and harmonic extraction is more popular for sub-THz frequency generation. [156] (530 GHz, 40 nm CMOS) presents a sub-THz radiator based on an injection-locked oscillator. The VCO comprises two 3-stage triple-push oscillators to form a 6-stage triple-push oscillator that benefits from layout constraint reduction, signal balance improvement, and output power enhancement. Its injection-locked push-push VCO precedes its triple-push VCO, which employs third harmonic extraction and achieves high loop gain at the frequency close to the $f_{max}$ of a transistor. Even simple designs can provide sub-THz oscillation, such as a voltage-controlled ring oscillator leveraging a power-combining network presented in [157] (363 GHz, 55 nm BiCMOS).

### H. UTILIZING RADIATOR ARRAYS

The sub-THz frequency generation is also leveraged for radiator array design [158]. [159] (460 GHz, 65 nm CMOS) presented a radiator array as a fully integrated coherent source based on coupled standing wave oscillator cells that uses a varactor-less tuning method, thus reducing additional loss and parasitics. Similarly to the prior art, [160] (450 GHz, 65 nm CMOS) uses coupled $2^{nd}$ harmonics to make a 2D scalable oscillator array. Each oscillator consists of a transis-tor with a T-embedding network, and the array is composed of differential oscillators along with high impedance quarter-wavelength resonators coupled adjacently in each row. This results in preventing $2^{nd}$ harmonic leakage from happening from the drain to the lossy gate and improves harmonic extraction efficiency. Aiming for a highly efficient THz radiator, [161] (318 GHz, 130 nm SiGe BiCMOS) proposes a harmonic VCO employing Colpitts structure in which adding lossless networks at the CM nodes results in tuning harmonic impedance and subsequently, changing the $2^{nd}$ harmonic current. Similarly, [162] employs a differential Colpitts configuration to produce a 64-pixel THz source. This configuration includes a doubler where the second harmonic, generated at the emitter, is fed reactively back to the base-emitter junction, thereby enhancing the second harmonic current. Similarly, [163] (308 to 317 GHz, 130 nm BiCMOS) uses VCOs for a source array. The VCO utilizes the differential Colpitts structure employing TLs and leveraging a cascode structure to provide low impedances while extracting the second harmonic. In addition, it uses a series-LC resonator at the base of core bipolar devices to provide RF grounding. In [164] (425 GHz, 14.6% TR, 90 nm BiCMOS), an intercoupled 2×3 array of PIN-diode-based Colpitts radiators operating in the fundamental frequency range of 80 to 92 GHz is used to realize the $5^{th}$ harmonic radiator with the output power of 18.1 dBm at 425 GHz. In [165] (469 to 489 GHz, 65 nm CMOS), a design methodology for 2D ring-coupled oscillator arrays is presented. Additionally, a 4×4 array prototype including a patch antenna and a quartz superstrate is measured, showcasing a high radiation efficiency per unit area. Similarly, [166] (643 to 689 GHz, 65 nm CMOS) presents a high-efficiency oscillator for third harmonic generation and extraction. In the presented coherent THz source array, the coherent oscillation mode radiates the third harmonic for output frequencies beyond 600 GHz without using a lens. In each unit cell, the gate is connected to the horizontal coupling network for out-of-phase oscillation, while the source is connected to the vertical coupling network for in-phase oscillation.

Pushing toward THz, [150] (0.68 to 0.72 THz, 65 nm CMOS) presents a 2D radiator array beyond 600 GHz for the first time. It uses $3^{rd}$ harmonic extraction from a coupled harmonic oscillator array with a fundamental frequency of 235 GHz. In each unit cell, two differential oscillators are coupled in phase. Moreover, each unit cell is horizontally and vertically coupled out-of-phase and in-phase, respectively, with adjacent cells at the fundamental frequency. In the array, the top and bottom boundaries are equivalent to perfect magnetic conductors, and the left and right boundaries are equivalent to perfect electric conductors. [168] (534 to 562 GHz, 130 nm SiGe BiCMOS) presents a triple-push oscillator with a common-collector topology. The traveling-wave oscillation mode, featuring a progressive 120-degree phase shift, is achieved by sharing the common-base output of each Colpitts stage through a power splitter network. This configuration causes the third harmonic currents generated at the base-emitter junctions of each unit to combine constructively at the output. As a result, the common-base output becomes AC-grounded at the fundamental frequency, effectively preventing frequency pulling. To isolate the oscillator from the bias network, base biasing is implemented using a quarter-wave stub terminated with a self-resonant capacitor tuned to the third harmonic.

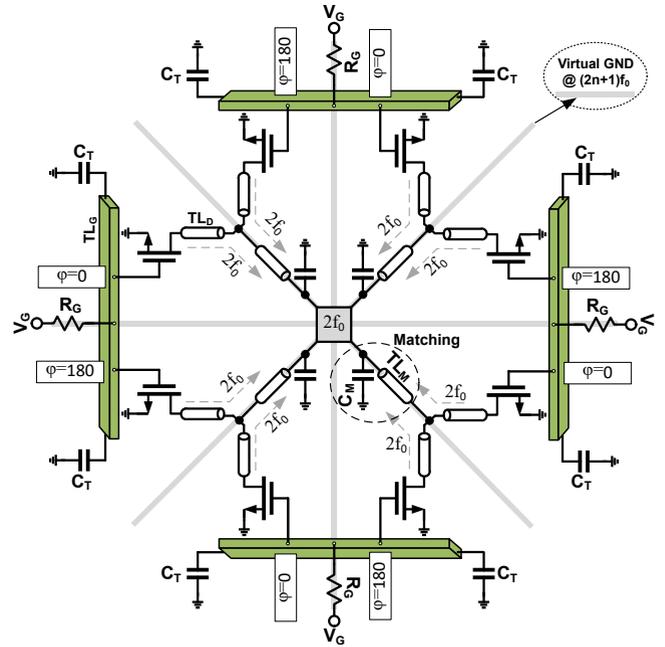

**FIGURE 9.** A high-power and wideband coupled standing wave VCO [167]. The oscillation relies on out-of-phase operation of the coupled SWOs, which creates virtual grounds at the $TL_D$ nodes, yielding inductive drain impedance and negative gate resistance. The phase relations are enforced by standing waves on the $TL_G$ and by $TL_D$ coupling between SWOs, causing the fundamental frequency and all odd harmonics to cancel and all symmetry planes to act as virtual grounds.

### I. OSCILLATORS IN NON-SILICON PROCESS TECHNOLOGIES

Non-silicon-based processes are also used to implement oscillators at higher frequencies. In [169] (180 to 184.4 GHz, 60 nm GaN), a degenerated common-source topology employing coplanar waveguide (CPW) structures assisted by some in-house device modeling is used to implement three different oscillators at the W band and the G band. The implemented VCO was the first reported oscillator at the G band in a GaN process at the time of publication. In [170] (92 to 103 GHz, 60 nm GaN HEMT), two frequency doublers and a buffer amplifier stage are cascaded to build a frequency quadrupler that transfers the signal from a VCO operating around 25 GHz to reach 100 GHz. The design achieved the best far carrier PN at the W band to date. In [171] (628 to 682 GHz, 250 nm InP HBT), a common-base cross-coupled push-push topology employing a coupled-line resonator for improved output power and efficiency in a compact size is presented. Three separate versions are implemented, achieving oscillation frequencies ranging from 509 to 682 GHz.

### J. NON-TRIVIAL APPROACHES

Non-trivial approaches for sub-THz frequency generation are not limited to large signal analysis. For standing wave VCOs (SWO), harmonic oscillators can generate close or beyond $f_{max}$ oscillations.

In [167] (219 to 238 GHz, 65 nm CMOS), a harmonic coupled SWO is presented in which four unit SWOs are coupled via drain TLs (i.e., $TL_D$) that simultaneously serve as drain inductors, inter-stage couplers, and second-harmonic combiners as illustrated in Fig. 9. This multifunctional TL network minimizes the extra loss typical of conventional coupling schemes while enforcing out-of-phase oscillation at the fundamental frequency between adjacent SWOs in order to start and maintain the oscillation, creating virtual grounds at all symmetry planes for the fundamental and its odd harmonics. This configuration ensures that even harmonics, particularly the second harmonic, are generated in-phase and efficiently routed and combined through $TL_D$ as well as the final $TL_M$ section, which also contributes to output matching.

In [172] (220 to 1140 GHz, 90 nm SiGe BiCMOS), the reverse recovery characteristic of p-i-n diodes is utilized to generate extremely narrow pulses with a repetition rate of up to 15 GHz. The reported equivalent isotropic radiated power (EIRP) levels in this work are among the highest to date. In [173] (110 GHz, TR 1.6%), the conditions for achieving maximum output power efficiency in the active devices of the VCO are determined through large-signal I-V analysis. The passive devices connected to the transistor terminals are then optimized based on these conditions, accounting for their losses and parasitic components. Furthermore, the parameters are calculated to ensure maximum power is delivered to the load while minimizing the power transferred to the subsequent frequency divider. In [174] (586.7 GHz, TR 0.7%, 40 nm CMOS), a mathematical modeling of the voltage-current relationship in the transistor of a high-frequency harmonic oscillator at the harmonic frequency is proposed. This modeling method facilitates the design of the second harmonic embedding network around the transistor, thus helping to propose a harmonic oscillator topology in which the requirement on the second harmonic voltages for fourth harmonic boosting is fulfilled. In an attempt to achieve higher oscillation frequencies with an older process technology [175] (241 to 251 GHz, 65 nm CMOS) proposed a capacitive load reduction circuit. This is achieved through an inductor placed between the gates of the differential output buffer that is coupled to the inductor of the core resonator. A second-harmonic push-push oscillator based on the same idea is also reported, operating at 432 GHz. In [176] (302 to 332 GHz, 130 nm BiCMOS), on-chip frequency stabilization is chosen over more common PLL stabilization, eliminating the need for dividers and off-chip crystal oscillators, and leading to lower power consumption and system integration cost. However, the proposed method suffers from a relatively high-temperature drift due to the temperature dependence of the dielectric constant of the back-end of line (BEOL) insulating layers used to implement the passive frequency sensing structure.

In summary, for sub-THz/THz oscillators—particularly at frequencies above 200 GHz—robust oscillation is achieved through a combination of design-specific optimization together with a set of common techniques that collectively ensure sufficient loop gain under marginal device speed conditions. For these oscillators, where the device $f_{max}$ is often only 2-3× higher than the oscillation frequency, achieving sufficient loop gain requires a combination of circuit-, device-, and layout-level optimization. Strong positive-feedback architectures, such as cross-coupled or transformer-based cores, are commonly employed to maximize effective negative resistance while minimizing excess phase shift. Capacitive or inductive neutralization is frequently used to partially cancel gate-drain or base–collector parasitic feedback, thereby improving unilateral gain near $f_{max}$. Device sizing is carefully optimized to balance transconductance against parasitic capacitances and series resistance, often favoring minimum-length devices with moderate width. High-frequency resonators and loads are co-designed to maximize effective impedance and quality factor despite increased passive losses at sub-THz frequencies. Biasing is selected to operate devices near peak $f_T$ and $f_{max}$, particularly in SiGe technologies, while extremely compact and symmetric EM-aware layouts are essential to prevent parasitic degradation of loop gain. Finally, many designs exploit harmonic operation to relax the fundamental oscillation frequency relative to $f_{max}$, enabling robust startup and stable oscillation under marginal device speed conditions.

## VI. CONCLUSION

This article presents recent advancements in mm-wave and sub-THz/THz oscillators for 5G/6G and beyond, where they play a pivotal role in enabling next-generation high-frequency communication systems, wireline/wireless technologies, and radar applications. It reviews the strengths and limitations of various state-of-the-art designs, categorizing mm-wave harmonic oscillators below 100 GHz into two main types: fundamental tone and harmonic tone, while also addressing sub-THz/THz oscillators operating above 100 GHz. Across the reviewed mm-wave fundamental-tone oscillators, the recurring design tension is the PN–TR–power triangle. Techniques that push phase noise down, such as high-Q LC tanks, transformer/series-resonant resonators, class-C/F waveform shaping, or multi-core coupling, tend to reduce frequency agility (or make tuning less linear), increase sensitivity to parasitics, and/or demand higher voltage swing and bias current to guarantee startup and robustness across PVT corners. Conversely, approaches that broaden tuning (switched-capacitor/varactor banks, multi-mode cores, reconfigurable coupling networks) often degrade effective tank Q through switch resistance, routing capacitance, and mode-dependent losses, which shows up as worse PN or higher power for the same noise target. Multi-core and quadrature solutions add another trade-off: they can improve phase noise and enable wideband I/Q generation, but they also introduce mode competition, spurs, and mismatch-induced penalties, making layout symmetry, coupling strength, and synchronization paths first-order constraints rather than *implementation details*. For the harmonic-tone and sub-THz/THz oscillators, the dominant trade-offs shift toward frequency reach versus efficiency and spectral purity. Getting above 100 GHz frequently leans on harmonic extraction, injection-locked multipliers, push-push/triple-push symmetry, distributed/standing-wave structures, or power-combining methods that relax device $f_{max}$ limits but typically incur PN multiplication, lower DC-to-RF efficiency, and strong sensitivity to amplitude/phase mismatch and EM parasitics that corrupt harmonic cancellation and raise unwanted tones. At these frequencies, interconnect/radiation losses, load pulling from buffers/antennas, temperature drift, and packaging/measurement interfaces become inseparable from the oscillator itself, so oscillator design becomes a co-design problem. Open problems that emerge across the surveyed works include: achieving simultaneously wide TR and low PN without switch/parasitic penalties; robust spur/mode control in many-core and symmetric harmonic architectures; better device/passive/noise modeling near $f_{max}$ (including temperature and aging) to reduce over-design; and practical co-optimization of oscillator–multiplier–radiator/package to manage efficiency, EMI, and frequency stability in real systems.

In this paper, key recent innovations are discussed, including bandwidth extension techniques, flicker noise reduction strategies, and PN improvement methods. Enabling significant improvement in data transfer rates, resolution, and sensing capabilities while embracing underlying challenges such as PN, power efficiency, and thermal management, it gives insight for better comparison and thoroughness, while ongoing innovations in material, design techniques, and integration continue to enhance the performance and reliability of high-frequency oscillators.

## ACKNOWLEDGMENT

The authors would like to thank Prof. Pietro Andreani for his fruitful insights and comments.